\documentclass[12pt]{article}
\usepackage[
    top=1in,
    bottom=1in,
    left=1in,
    right=1in
]{geometry}
\usepackage[utf8]{inputenc}
\usepackage{amsmath, amssymb, amsthm}
\usepackage{graphicx, float}
\usepackage{svg}
\usepackage{enumitem}
\usepackage{xcolor}
\usepackage{siunitx}
\usepackage[version=4]{mhchem}
\usepackage{booktabs, tabularx, threeparttable}
\usepackage{longtable}
\usepackage{array}
\usepackage{microtype}
\usepackage{soul}
\usepackage{etoolbox}
\usepackage{chngcntr}
\usepackage{titlesec}
\usepackage{graphicx}
\usepackage{placeins}
\usepackage{float}
\usepackage{caption}
\usepackage{url}
\usepackage{doi}
\usepackage{cite}
\usepackage{tabularx}
\usepackage{caption}
\definecolor{darkblue}{RGB}{0,51,153}
\definecolor{DarkPink}{RGB}{231,84,128}
\definecolor{HotPink}{RGB}{255,105,180}
\definecolor{DeepPink}{RGB}{255,20,147}
\definecolor{BlueSentence}{RGB}{0,80,180}
\definecolor{RedSentence}{RGB}{180,0,0}

\titleformat{\section}{\normalfont\bfseries}{\thesection}{1em}{}
\titleformat{\subsection}{\normalfont}{\thesubsection}{1em}{}
\titleformat{\subsubsection}{\normalfont}{\thesubsubsection}{1em}{}

\usepackage{hyperref}
\hypersetup{
    colorlinks=true,
    linkcolor=darkblue,
    citecolor=darkblue,
    urlcolor=darkblue,
    pdftitle={Fine Oxide Dispersoids Modulate Phonon Drag and Dislocation Relaxation in Dynamically Deformed Superalloys},
    pdfauthor={Naveen Dinujaya et al.}
}

\newpage
\begin{document}

\begin{flushleft}
\Large \textbf{Fine Oxide Dispersoids Modulate Phonon Drag and Dislocation Relaxation in Dynamically Deformed Superalloys}
\end{flushleft}

\begin{flushleft}
\normalsize
Naveen Dinujaya$^{1,2}$, 
Suhas Eswarappa Prameela$^{*,1,2,3}$
\end{flushleft}

\vspace{0.5em}

{\small
\noindent $^{1}$ Department of Materials Science and Engineering, University of Utah, Salt Lake City, UT, USA \\
\noindent $^{2}$ Department of Metallurgical Engineering, University of Utah, Salt Lake City, UT, USA \\
\noindent $^{3}$ Department of Mechanical Engineering, University of Utah, Salt Lake City, UT, USA \\

\noindent *Corresponding author: suhas.prameela\,@\,utah.edu

\vspace{2em}
\hrule

\begin{abstract}
\noindent Plastic deformation in metallic alloys is primarily governed by the motion of dislocations, which are atomic scale line defects that move through the crystal lattice under an applied stress. Superalloys containing fine oxide particles withstand extreme temperatures for prolonged durations by resisting dislocation motion. This thermally controlled mechanism conventionally involves dislocations first climbing over the particle and slowly detaching from it, a process that typically occurs on the order of seconds. However, these mechanisms drastically change when dislocation velocities increase, potentially exceeding half the shear wave speed of the material. At such extreme speeds, dislocations can interact with lattice vibrations, leading to pronounced phonon interactions. We leverage a pulsed laser to drive rigid  microspheres at controlled velocities towards superalloy substrates containing a dense oxide dispersion. Synchronized high speed imaging allows precise mapping of deformation events, allowing high throughput decoupling and modeling of plasticity contributions. We find that the oxide network produces a dual and counterintuitive effect. Our modeling framework indicates that rapidly moving dislocations bypass oxide particles by bowing rather than climbing, thereby suppressing departure side dislocation relaxation. At the same time, the dense oxide network confines fast moving dislocations within the critical interparticle distance, thereby reducing their interaction with phonons. These findings shed light on new plasticity mechanisms in oxide particle containing superalloys when line defects accelerate and dissipate energy on picosecond timescales.
\end{abstract}

\hrule


\newpage

\section{Introduction}

\noindent Imperfections in crystalline solids govern the mechanical properties and performance of metals~\cite{eswarappa2020defect,cai2016imperfections}. Among them, dislocations are line defects within the crystal lattice that play a central role in plastic deformation~\cite{breithaupt2023dislocation,hirth2025revised}. The collective movement of these dislocations under applied stress is responsible for the macroscopic plasticity of metals~\cite{chen2020unusual}. Therefore, controlling defect motion is central to strengthening metals and requires modulating how dislocations move, interact with, and bypass obstacles. This is especially important for superalloys, which are used in propulsion, power generation, and nuclear applications~\cite{gradl2024extreme,peters2024materials,jia2022first,pollock2006nickel,alay2023tailoring}. These materials must retain strength while limiting plastic deformation under extreme conditions ~\cite{smith20233d,zhang2022inhibiting}. Traditionally, deformation in superalloys is often viewed as a gradual process driven by slow dislocation motion. Over these longer time scales, dislocations can undergo thermally activated processes such as climb~\cite{arzt1986threshold}, detachment~\cite{rosler1990new}, and interfacial dislocation relaxation~\cite{arzt1988kinetics}, thereby influencing plastic flow.\\

\noindent Dynamic deformation introduces a different set of constraints on defect motion. When metals deform rapidly, dislocations can accelerate to velocities exceeding half the material's shear wave speed before conventional thermally activated processes can operate~\cite{tang2024quantifying,wu2026smaller}. In this regime, plastic deformation is governed not only by the obstacles that dislocations encounter but also by their interactions with the vibrating lattice. Fast moving dislocations exchange energy and momentum with the lattice through phonons~\cite{marian2006moving,li2020quantized}. At high dislocation velocities, interactions with lattice vibrations dissipate energy and contribute to flow resistance through phonon drag~\cite{dowding2024metals}. Although phonon drag is recognized as an important contribution to flow strength under dynamic loading, how dispersed oxide particles influence phonon-mediated dislocation motion in superalloys remains largely unexplored.\\

\noindent Quantifying phonon drag is especially important because oxide particles introduce competing spatial and temporal scales for fast dislocation motion~\cite{zhang2023evolution,eswarappa2024high,borasi2025self}. Under slow deformation, a dense oxide network strengthens the metal by forcing dislocations to climb and detach, thereby acting as a dislocation relaxation mechanism~\cite{kassner2015fundamentals}. However, under rapid deformation, the same dense oxide network may confine the dislocation glide before phonon drag becomes a substantial contribution to flow strength~\cite{tang2025suppressed,wu2026smaller}. Thus, the dense oxide network may not simply resist dislocation glide but may also regulate the distance over which dislocations can accelerate and, consequently, the energy dissipated through interactions with the vibrating lattice. This interplay suggests that, under dynamic loading, oxide particles may regulate not only dislocation motion but also the development of phonon drag mediated resistance. \\

\noindent In this study, we utilized controlled microparticle impact testing to probe the dynamic regime by measuring impact and rebound events with synchronized high speed imaging at millions of frames per second. This testing approach enables localized deformation at characteristic strain rates approaching $10^7 \text{ s}^{-1}$ across a range of temperatures, thereby linking the changes in dynamic flow strength to the underlying defect structures and kinetics. Our results reveal that dense oxide networks act as kinetic regulators of dynamic deformation in superalloys rather than passive dislocation barriers. Specifically, our framework indicates that the oxide network restricts dislocation acceleration and the time available for dislocation relaxation at particle interfaces.

\section{Results}

\noindent The dynamic flow strength was determined using a three step process. First, microparticles were launched using the laser driven microparticle impact system described in the methods section and illustrated in Fig.~\ref{Fig1}A. Second, the impact and rebound events were recorded using high speed imaging to determine the corresponding particle velocities. Third, these measured velocities were used to calculate the coefficient of restitution and extract the dynamic flow strength of the superalloy. GRX-810~\cite{smith20233d} was selected as a model superalloy because it contains fine hexagonal shaped yttria particles, as shown in Fig.~\ref{Fig1}B and C. These oxide particles have an average major diameter of 25 nm, as depicted in Fig.~\ref{Fig1}D and confirmed by multi-particle image analysis (SI Appendix, Fig.~S1A and B). The energy dispersive X-ray spectroscopy (EDS) elemental maps in Fig.~\ref{Fig1}E show that they are enriched in yttrium (Y).

\begin{figure}[H]
    \centering
    \includegraphics[width=\linewidth]{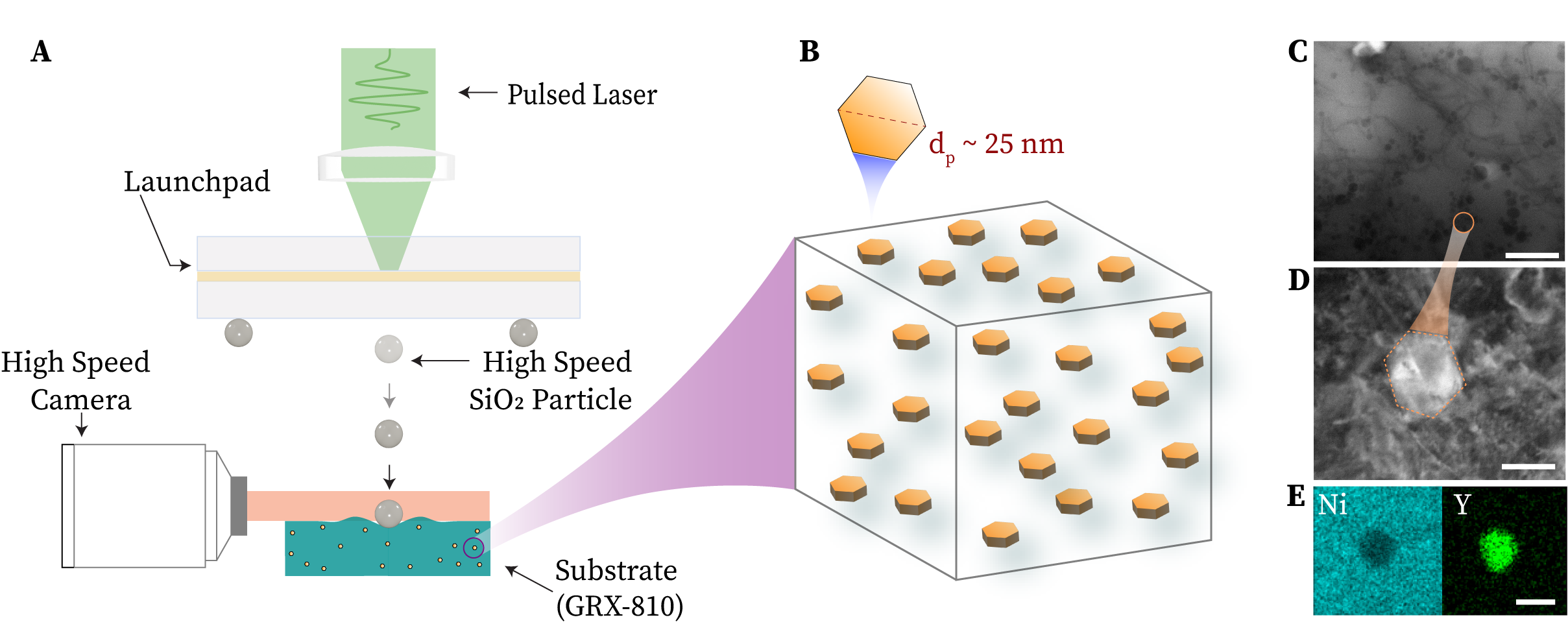}
  \caption{\small \textbf{High throughput microparticle impact testing and nanoscale oxide structure of the superalloy substrate.}
(A) Schematic of the laser driven microparticle impact system used to probe the localized dynamic deformation.
(B) Schematic diagram illustrating the distribution of oxide particles in the superalloy. The particles exhibit an average major diameter of 25 nm.
(C) High magnification STEM HAADF image of the dense oxide network. (D) High resolution view of a fine oxide particle, accompanied by (E) representative EDS elemental maps showing the Ni matrix and Y enrichment within the particle. Scale bars are 200 nm for (C) and 25 nm for (D, E).
}
\label{Fig1}
\end{figure}

\noindent The contrasting dynamic response of the oxide particle containing superalloy was captured using synchronized high speed imaging at four temperatures. Tests were conducted at $20~^\circ\mathrm{C}$, $100~^\circ\mathrm{C}$, $155~^\circ\mathrm{C}$, and $204~^\circ\mathrm{C}$ using a $\sim 14~\mu\mathrm{m}$ diameter silica microparticle. Representative in situ impact and rebound snapshots for all four test conditions are shown in Fig.~\ref{Fig2}A for an impact velocity of $\sim143~\mathrm{m/s}$. The corresponding videos are provided in SI Appendix, Movies S1-S4. In each case, the silica microparticle retained its spherical shape after impact, with no visible evidence of shattering. \\

\noindent At all test temperatures, the measured rebound velocity was lower than the impact velocity, indicating irreversible kinetic energy loss during impact. As the substrate temperature increased, the rebound velocity decreased at comparable impact velocities, indicating greater impact energy dissipation at elevated temperature. To evaluate the dependence of kinetic energy dissipation on impact velocity, the coefficient of restitution (CoR)~\cite{10.1115/1.2787319} is calculated as

\begin{equation}
\mathrm{CoR} = \frac{v_r}{v_i},
\end{equation}

\noindent where $v_r$ and $v_i$ are the rebound and impact velocities, respectively. For each temperature condition, $\mathrm{CoR}$ was plotted as a function of impact velocity on a double logarithmic scale. 
\vspace{0.3cm}

\begin{figure}[H]
    \centering
    \includegraphics[width=\linewidth]{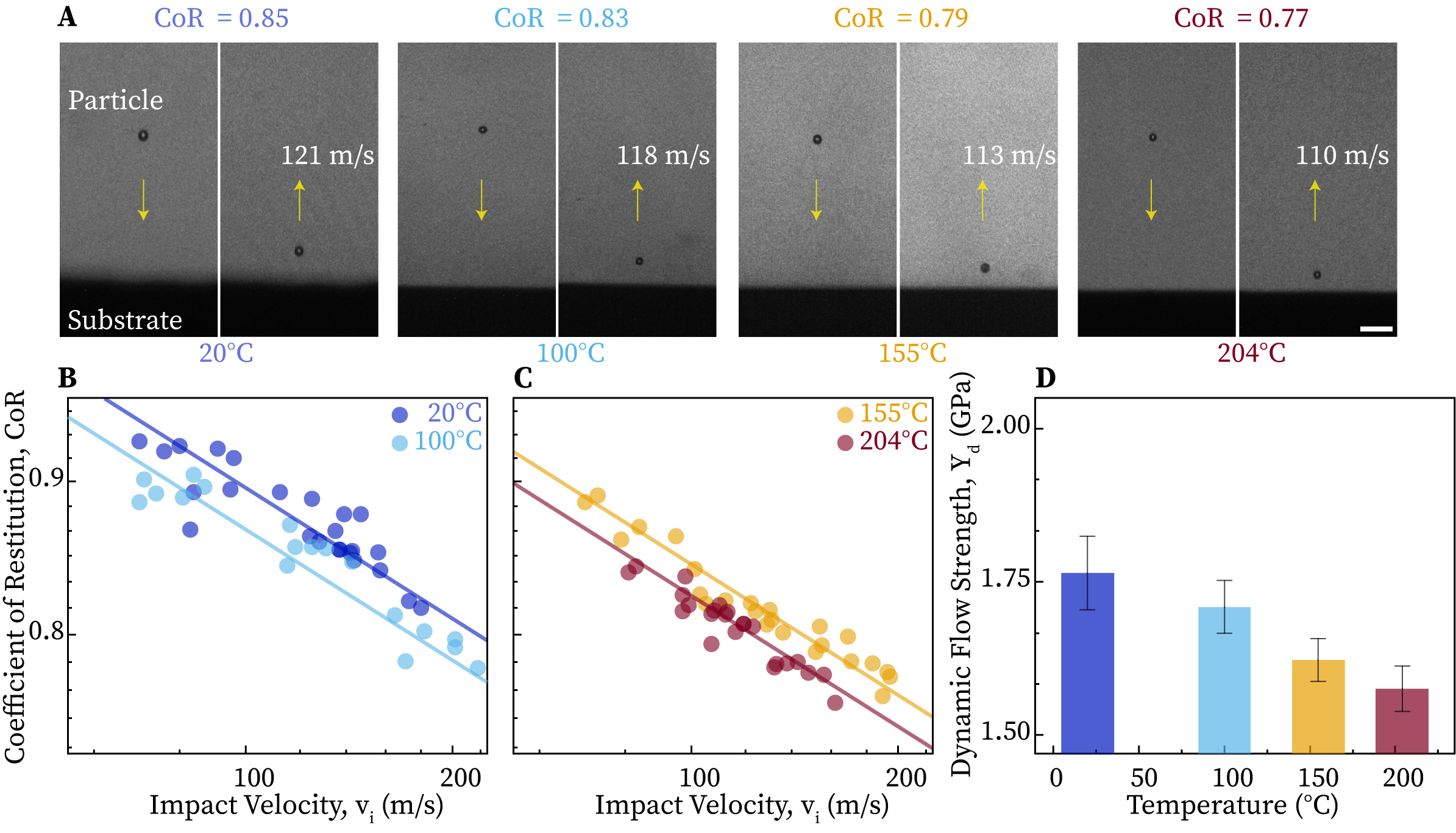}
  \caption{\small{\textbf{In situ microparticle impact dynamics and dynamic flow strength evaluation.}}
(A) High speed videography images showing representative impact and rebound events at an impact velocity of $v_i\approx 143~\mathrm{m/s}$ at temperatures of $20~^\circ\mathrm{C}$, $100~^\circ\mathrm{C}$, $155~^\circ\mathrm{C}$, and $204~^\circ\mathrm{C}$.
(B) and (C) $\mathrm{CoR}$ plotted as a function of impact velocity on a double logarithmic scale.
(D) Extracted dynamic flow strength as a function of test temperature. Scale bar for (A) is $50\ \mu\text{m}$.
}
\label{Fig2}
\end{figure}

\noindent Kinetic energy dissipation increases systematically with impact velocity across all temperature conditions, consistent with greater plastic deformation of the substrate at higher velocities. The corresponding CoR–velocity data for all test temperature conditions are shown in Fig.~\ref{Fig2}B and C, with this trend being a progressive decrease in CoR with increasing impact velocity. This behavior is characteristic of impact by an elastic sphere against a plastically deforming substrate~\cite{liu2024model}. To extract dynamic flow strength, the CoR–velocity data were fitted using the semi empirical framework of Wu et al.~\cite{wu2003rebound,wu2005energy}, which has also been widely adopted in prior microparticle impact studies~\cite{sun2020transition,liu2026modelling,lucas2025measuring}.

\begin{equation}
\mathrm{CoR}
=
\alpha
\left[
\left(
\frac{v_y}{v_i}
\right)
\left(
\frac{E^{*}}{Y_d}
\right)
\right]^{1/2}
\label{eqcor_wu}
\end{equation}

\noindent where $\alpha= 0.78$ is a prefactor corresponding to the impact of an elastic sphere~\cite{wu2005energy,liu2024model,lucas2025single}. In the present analysis, the silica microparticle is modeled as the elastic sphere, while the superalloy is treated as an elastic perfectly plastic substrate. The parameter $v_y$ is the yield velocity corresponding to the onset of plastic deformation (SI Appendix, Eq. S6), and $E^*$ is the effective elastic modulus of the contacting materials, accounting for both the silica microparticle and the substrate, as described in SI Appendix, Eq.~S7. The parameter $Y_d$ is the effective dynamic flow strength of the substrate and was inferred from the CoR-velocity response using Eq.~\ref{eqcor_wu}. In this work, $Y_d$ is interpreted as an event averaged strength that summarizes the substrate response during impact, rather than as a pointwise flow stress~\cite{lucas2024microstructure,johnson1970correlation,10.1115/1.2787319}.\\

\noindent The dynamic flow strength results reveal two important features of the superalloy's dynamic response. First, the oxide particle containing superalloy exhibits higher flow strength under dynamic loading than under quasi-static loading conditions, and this difference remains evident at elevated temperatures (compare Fig.~\ref{Fig2}D with SI Appendix, Fig.~S3D). The calculated dynamic flow strengths are $1764 \pm 60~\mathrm{MPa}$ at $20~^{\circ}\mathrm{C}$, $1709 \pm 43~\mathrm{MPa}$ at $100~^{\circ}\mathrm{C}$, $1623 \pm 35~\mathrm{MPa}$ at $155~^{\circ}\mathrm{C}$ and $1576 \pm 37~\mathrm{MPa}$ at $204~^{\circ}\mathrm{C}$, where the reported uncertainties represent one standard deviation. Second, the superalloy exhibits a decrease in dynamic flow strength with increasing temperature, consistent with temperature  dependent weakening of the governing strength contributions \cite{dowding2024metals,dowding2026extreme}. These findings are independently supported by experimentally measured dynamic hardness values, which exhibit an identical downward trajectory with temperature (SI Appendix, Fig.~S2E).

\section{Discussion}

\noindent To estimate the contribution of each strengthening mechanism, we decomposed the dynamic flow strength of the GRX-810 superalloy, as detailed in SI Appendix, Section 7.3 and Fig.~S4. This decomposition combines thermally activated dislocation motion, athermal microstructural resistance associated with dislocation interactions and boundaries, oxide particle resistance, and dynamic dislocation resistance arising from interactions between rapidly moving dislocations and lattice phonons. While prior microparticle impact studies have successfully utilized this decoupled mechanistic framework for various pure metals~\cite{tang2024quantifying,dowding2024metals,dowding2026extreme}, we have extended it here to oxide particle containing superalloys. These prior studies further indicate that the transition into the phonon drag regime is gradual rather than abrupt, such that several strengthening mechanisms remain active concurrently even under conditions of extremely fast dislocation motion. Probing how these contributions partition has recently uncovered counterintuitive dynamic behavior of metals~\cite{dowding2024metals}, while providing a systematic basis for isolating the additional resistance introduced by the oxide particle network relative to other concurrent mechanisms.

\subsection{Role of Phonon Drag During Fast Dislocation Motion}

\noindent Localized dynamic impact produces the high applied stresses that enable dislocations to attain high velocities (Fig. \ref{Fig 3}A and B). In this regime, dislocation motion is opposed by a velocity dependent phonon drag force. For the present microparticle test conditions and elevated temperature range, this resistance is primarily due to interactions between moving dislocations and phonons in the lattice. Consequently, this phonon drag strength is expected to contribute significantly to the dynamic flow strength, particularly as temperature increases the phonon drag coefficient~\cite{dowding2024metals}. Contrary to this expectation, flow strength decomposition (SI Appendix, Section 7.3) reveals that the phonon drag contribution in the present superalloy is significantly reduced. This discrepancy can be explained by the nanometer scale barriers introduced by the dense oxide particle network, which confine dislocation motion over much shorter distances (Fig.~\ref{Fig 3}B). Under these conditions, the oxide network may restrict dislocation motion, limiting the development of strong dislocation–phonon interactions. Using this framework, the phonon drag contribution was fitted using a scaling factor (SI Appendix, Eq.~S28), yielding an approximately threefold reduction relative to the absence of the oxide network (Fig.~\ref{Fig 3}C). This fitted reduction can be assessed more directly by estimating how nanometer scale confinement limits the development of dislocation phonon interactions, thereby providing a quantitative framework for the reduced phonon drag contribution inferred from the present experiments.\\

\noindent The dislocation confinement interpretation can be examined more directly using the treatment of Gillis and Kratochvil~\cite{gillis1970dislocation}. This approach models the acceleration of a single dislocation under an applied shear stress in the presence of a resisting drag force.
This treatment was later adapted by Tang et al.~\cite{tang2025suppressed} to develop a framework for microparticle impact experiments. 
Within this approach, the dislocation acceleration within the superalloy is expressed as

\begin{equation}
\label{eqdislocation_accel}
\frac{dv}{dt} = \frac{\tau_{AP}\left(\Gamma^{\frac{3}{2}}\right)}{b\rho_s} - \frac{B_d(T) v\left(\Gamma^{\frac{1}{2}}\right)}{b^2 \rho_s}
\end{equation}

\noindent where $\frac{dv}{dt}$ is the dislocation acceleration, $\tau_{\mathrm{AP}}$ is the applied shear stress (SI Appendix, Section 8), $B_d(T)$ is the drag coefficient, $\rho_s$ is the mass density of the superalloy, and $b$ is the Burgers vector. The relativistic factor is defined as $\Gamma = ({1 - \frac{v^2}{c^2}})$, in which $v$ and $c$ are the dislocation velocity and shear wave velocity, respectively. To evaluate how accelerating dislocations behave in this superalloy, we examine the terminal limit at which acceleration drops to zero and the dislocation reaches steady state. At this point, the external driving force perfectly balances the internal lattice vibration resistance.\\

\noindent To express this balance in dimensionless form, the dislocation velocity is normalized by the material's shear wave velocity. This normalization reduces the equation to a competition between the normalized driving stress term ($D^*$) and the normalized drag resistance term ($R^*$) described in SI Appendix, Section 9. This competition is quantified by the drag participation factor

\begin{equation}
\Phi_D
=
\frac{R^*}{D^*}
=
\frac{B_d(T)v}
{\tau_{\mathrm{AP}}b\Gamma}
\label{eqdrag_participation}
\end{equation}

\noindent The drag participation factor represents the ratio of the drag resistance to the driving stress. A value of $\Phi_D<1$ indicates that the driving stress remains dominant and the dislocation continues to accelerate. The theoretical steady state condition occurs when $\Phi_D=1$, corresponding to an exact balance between the driving and drag terms, as shown in Fig.~\ref{Fig 3}D. Equation~\ref{eqdislocation_accel} was recast using $dx/dv=v/(dv/dt)$ and numerically integrated to obtain the dislocation velocity and acceleration as functions of distance traveled within the superalloy matrix (SI Appendix,
Section 10). Thermally activated strength contributes to the overall flow strength, while Fig.~\ref{Fig 3}E illustrates the increasing role of phonon drag as dislocations accelerate between obstacles. Dislocations accelerate rapidly over short distances, but the rate of acceleration progressively decreases as the drag participation factor approaches unity. Consequently, the dislocation velocity approaches the theoretical steady state value asymptotically. Using the present matrix properties for $20~^\circ\mathrm{C}$, the model predicts a theoretical steady state dislocation velocity of approximately $0.725c$ (Fig.~\ref{Fig 3}D). With increasing temperature, this balance shifts progressively toward lower steady state velocities of $0.651c$ at $100~^{\circ}\mathrm{C}$, $0.564c$ at $155~^{\circ}\mathrm{C}$, and $0.506c$ at $204~^{\circ}\mathrm{C}$ (SI Appendix, Fig.~S5D). The exact steady state velocity is approached asymptotically and therefore cannot be reached over a finite distance. Accordingly, 99.5\% of steady state velocity is used as the numerical convergence criterion. Thus, $\gamma$ denotes the uninterrupted glide distance required to reach this criterion. The predicted steady state velocities align with prior studies reporting dislocation velocities between $0.6c$ and $0.8c$ \cite{bryukhanov2020dynamics}. 

\begin{figure}[H]
    \centering
    \includegraphics[width=\linewidth]{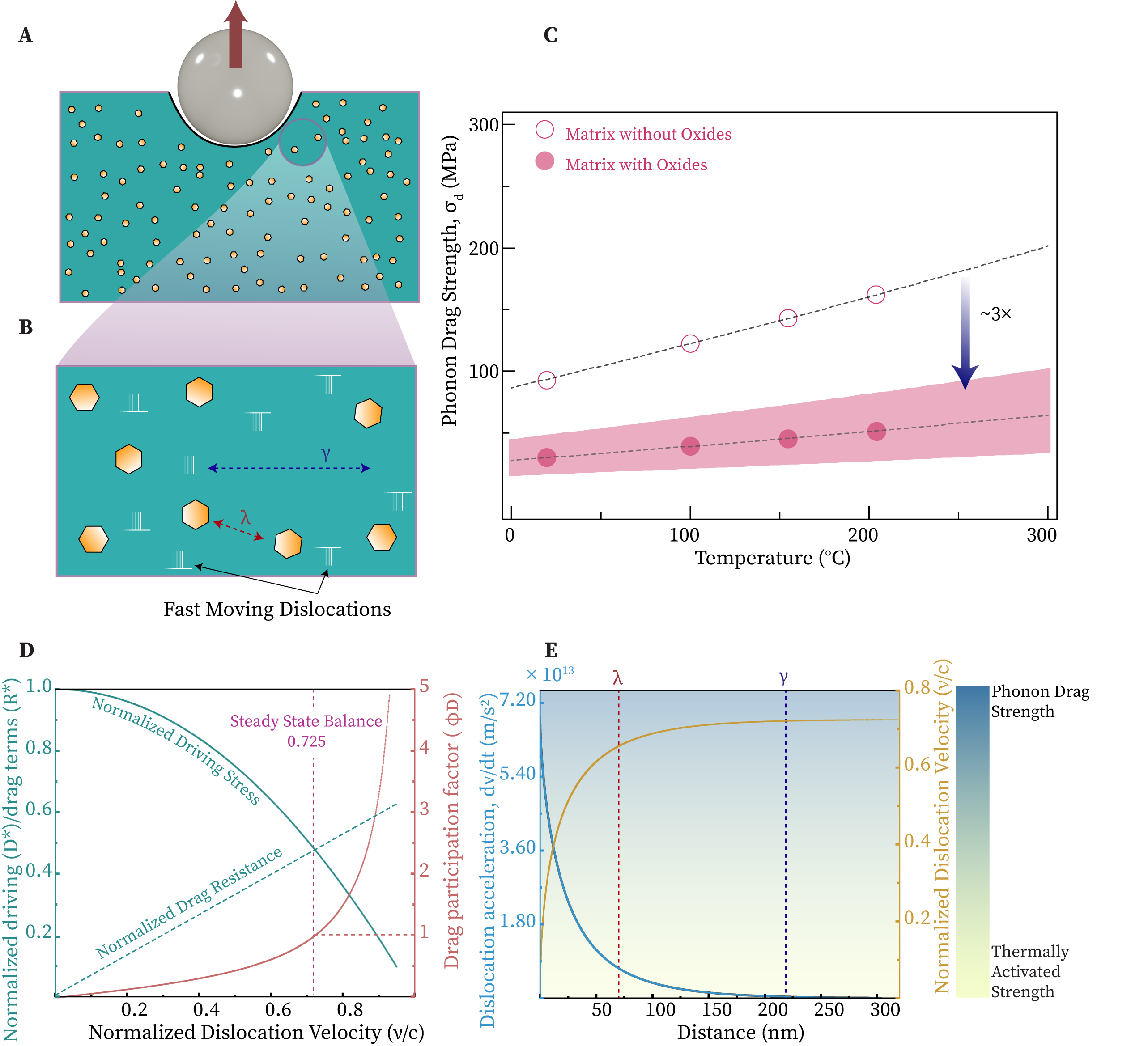}
  \caption{\small\textbf{Phonon drag kinetics during fast dislocation motion.}
(A) Schematic illustration of microparticle impact dynamics generating extreme dynamic deformation. 
(B) Magnified schematic of the substrate matrix showing fast moving dislocations interacting with oxide dispersoids, defined by interparticle spacing ($\lambda$) and uninterrupted dislocation glide distance ($\gamma$). 
(C) Temperature dependence of phonon drag strength for the matrix without oxides (open circles) and matrix with oxides (filled circles), highlighting an approximate $3\times$ reduction in drag strength in the presence of oxide dispersoids. The band represents the lowest ($5.07 \times 10^6~\mathrm{s}^{-1}$) and highest strain rates ($1.58 \times 10^7~\mathrm{s}^{-1}$) achieved in test conditions for the matrix with oxides, and a dashed line represents a strain rate of $10^7~\mathrm{s}^{-1}$.
(D) Normalized driving stress ($D^*$) and drag resistance terms ($R^*$) as a function of normalized dislocation velocity, showing the evolution of normalized stress, drag resistance, and drag participation factor at $20~^\circ\mathrm{C}$, with steady state balance achieved at $v/c \approx 0.725$. 
(E) Dislocation acceleration and normalized velocity as a function of distance, overlaid with the interparticle spacing and uninterrupted dislocation glide distance at $20~^\circ\mathrm{C}$, where the glide distance corresponds to reaching 99.5\% of the steady state velocity, illustrating the transition from thermally activated strength to phonon drag strength.}

\label{Fig 3}
\end{figure}

\noindent These values are physically reasonable for pure Ni and related Ni alloys \cite{olmsted2005atomistic,zhao2017atomic}, which share similar behavior with the tested superalloy.\\

\noindent The role of oxide confinement can be understood by considering an otherwise comparable superalloy without oxide particles, in which dislocations would travel over longer uninterrupted distances and develop a greater phonon drag force. For pure Ni and related Ni alloys, velocities of this magnitude are sufficiently high for dislocation phonon interactions to become significant~\cite{olmsted2005atomistic,zhao2017atomic,anis2024dislocation}. Under dynamic deformation, continued acceleration increases the exchange of energy and momentum between moving dislocations and lattice vibrations. However, closely spaced oxide particles in the superalloy interrupt this acceleration process. Therefore, the relevant mechanistic limitation is not a sharp velocity threshold but the restricted glide distance imposed by the dense oxide network. By repeatedly shortening the available path, the oxide particles limit the phonon drag contribution reached during dislocation motion.\\

\noindent TEM observations and theoretical calculations (SI Appendix, Eq.~S2) reveal a dense oxide particle network within the GRX-810 superalloy with a particle number density of $2.20 \times 10^{21}\ \text{m}^{-3}$ (Fig.~S1A). This high particle density significantly shortens the dislocation mean free path. Consequently, the effective interparticle spacing ($\lambda$) is restricted to $70\ \text{nm}$, which substantially constrains the distance over which dislocations can accelerate. By comparison, our distance based analysis using the dislocation acceleration model indicates that a glide distance ($\gamma$) of approximately $219\text{ nm}$ is required for the drag controlled response to reach 99.5\% of the steady state velocity. Thus, the presence of fine oxide particles reduces the available glide distance in the superalloy to approximately 32\% of the distance required without oxide dispersion. As a result, the phonon drag contribution to strengthening is reduced in the superalloy.\\

\noindent Although a short mean free path is necessary to suppress phonon drag, this structural constraint must also persist under the extreme stresses, high strain rates, and adiabatic heating generated during impact~\cite{tiamiyu2022nanotwinning}. Prior work confirms that the oxide particle network maintains its structure under extreme temperature conditions~\cite{heczko2026microstructural}. Consistent with limited coarsening over the investigated thermal exposure~\cite{de2021thermal,mao2023transformation}, the interparticle spacing is treated as unchanged during impact, ensuring that drag controlled dislocation motion cannot fully develop in the superalloy.

\subsection{Role of Dislocation Relaxation During Fast Dislocation Motion}
\label{relaxation}

\noindent Oxide dispersion strengthened superalloys derive their elevated temperature strength from fine oxide particles that remain stable at high temperatures. As dislocations approach the oxide particles during slow motion (Fig.~\ref{FIG 4}A, i), the particles impede glide, requiring bypass by climb (Fig.~\ref{FIG 4}A, ii). Rösler and Arzt~\cite{rosler1990new} described how a dislocation first climbs around an oxide particle but can remain strongly pinned at the departure side because of attractive particle-dislocation interactions. These interactions prolong dislocation pinning, shifting the rate controlling mechanism from dislocation climb to thermally activated dislocation detachment (Fig.~\ref{FIG 4}A, iii). Within this model, dislocation relaxation at the oxide–matrix interface is governed by a local reduction in dislocation line energy. This local energetic relaxation is quantified by the relaxation factor ($k$), which is defined as the ratio of the line energy of the dislocation segment attached at the departure side of the oxide particle to the reference line energy prior to interfacial relaxation. The relaxation factor inferred for GRX-810 ($k=0.77$) from prior data~\cite{semiatin2024constant,smith2025mechanisms} is lower than the values reported for similar alloys~\cite{rosler1990new}. This lower value indicates a greater reduction in dislocation line energy at the oxide–matrix interface during slow dislocation motion (SI Appendix, Section 7.2).
\\

\begin{figure}[H]
    \centering
    \includegraphics[width=\linewidth]{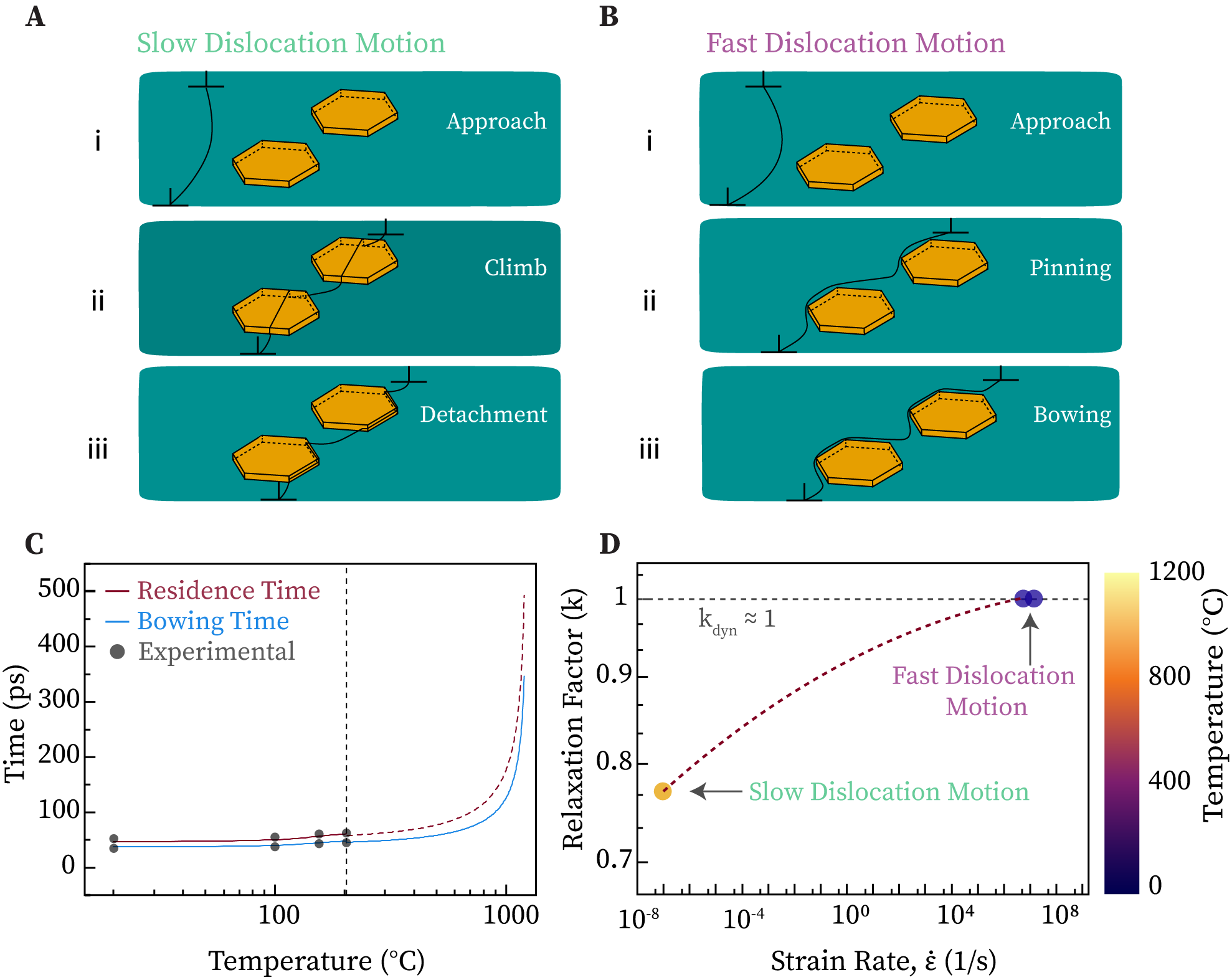}
  \caption{\small\textbf{Dislocation relaxation kinetics under fast motion.}
(A) Schematic of slow dislocation motion showing (i) approach, (ii) climb around oxide particles, and (iii) departure side detachment.
(B) Schematic of fast dislocation motion showing (i) approach, (ii) pinning, and (iii) bowing around oxide particles.
    (C) Comparison of dislocation bowing time and residence time across temperatures at a strain rate of $10^7~\mathrm{s}^{-1}$. Solid lines represent calculations using experimentally determined loading conditions, gray symbols denote calculated times at the four experimental temperatures, and dashed lines indicate calculations using model predicted loading conditions.
    (D) Strain rate dependence of the relaxation factor across temperatures, demonstrating that the relaxation factor is close to unity ($k_{\mathrm{dyn}}\approx1$) at extreme strain rates. The $k$ value for the lower strain rate (slow dislocation motion) is calculated from prior creep data \cite{semiatin2024constant,smith2025mechanisms}.}
\label{FIG 4}
 
\end{figure}

\noindent During fast dislocation motion, the available interaction time is expected to favor particle bypass by bowing rather than thermally activated climb. The schematic illustrates dislocation approach (Fig.~\ref{FIG 4}B, i), pinning at oxide particles (Fig.~\ref{FIG 4}B, ii), and subsequent bowing around the particles (Fig.~\ref{FIG 4}B, iii). This competition can be evaluated by comparing the dislocation residence time at the oxide–matrix interface with the characteristic times required for bowing and climb. From $20$ to $204~^\circ\mathrm{C}$, bowing and residence times were calculated using experimentally inferred dynamic flow strengths, with interpolation between the four test temperatures. Over this range, the calculated bowing time, $t_{\mathrm{bow}}$, increases from $37.6~\mathrm{ps}$ to $48.2~\mathrm{ps}$, whereas the residence time, $t_{\mathrm{res}}$, increases from $47.0~\mathrm{ps}$ to $61.2~\mathrm{ps}$. Above $204~^\circ\mathrm{C}$, model predicted dynamic flow strengths were used to extend the calculations to $1200~^\circ\mathrm{C}$, where $t_{\mathrm{bow}}$ and $t_{\mathrm{res}}$ reach approximately $346.7~\mathrm{ps}$ and $492.6~\mathrm{ps}$, respectively (Fig.~\ref{FIG 4}C). For corresponding temperature conditions, the residence time is 1.25 - 1.42 times the bowing time, indicating that the dislocation remains within the oxide interaction region long enough to bow around the particle. By contrast, the characteristic time required for thermally activated climb remains substantially longer than the available dislocation residence time across both the experimental and model predicted temperature ranges (SI Appendix, Section 12). Thus, climb assisted bypass is kinetically inaccessible during the extremely short dislocation-particle interaction. Accordingly, these estimates support that Orowan bowing~\cite{orowan1948symposium} represents the physically plausible oxide bypass mechanism under the present fast dislocation motion conditions.\\

\noindent According to dislocation residence time calculations and vacancy diffusion calculations (SI Appendix, Sections 11-13), dislocation relaxation on the departure side of the oxide particle is effectively suppressed during fast dislocation motion. To develop a relaxed departure side configuration, the dislocation must undergo climb assisted bypass. However, this mechanism is prevented by the extremely fast dislocation motion at high strain rates. The time required for climb assisted bypass is much longer than the dislocation residence time at the oxide–matrix interface. The extent of relaxation within this limited interaction time is quantified using the dynamic relaxation factor

\begin{equation}
k_{\mathrm{dyn}}
=
k+
(1-k)
\exp\left(
-\frac{t_{\mathrm{res}}}{\tau_{\mathrm{rel}}}
\right)
\label{eqdynamic_relaxation}
\end{equation}

\noindent where $k$ is the relaxation factor during slow dislocation motion, $t_{\mathrm{res}}$ is the dislocation residence time, and $\tau_{\mathrm{rel}}$ is the characteristic time required for vacancy mediated atomic redistribution. Since $t_{\mathrm{res}}\ll\tau_{\mathrm{rel}}$, the calculated dynamic relaxation factor approaches unity ($k_{\mathrm{dyn}}\approx1$) under extreme strain rate conditions, as depicted in Fig.~\ref{FIG 4}D and SI Appendix, Section 13. This behavior is expected to remain active even at elevated temperatures because the oxide particles are extremely thermally stable under these conditions~\cite{smith2025mechanisms,heczko2026microstructural}. Although increasing temperature may accelerate dislocation climb, the climb assisted bypass time still does not approach the extremely short dislocation residence time at the oxide–matrix interface. Therefore, departure side dislocation relaxation remains strongly suppressed even under combined extreme strain rate and elevated temperature conditions. 

\subsection{Interplay Between Oxide Dispersion and Phonon Drag Strengthening During Fast Dislocation Motion}

\noindent Strengthening from oxide particles is enhanced under fast dislocation motion compared with slower dislocation motion as shown in Fig.~\ref{figoxide_dispersion_phonon_drag}A. In this fast regime, the strengthening is most appropriately described by dislocation bowing between oxide particles, as shown in Fig.~\ref{FIG 4}B. Climb assisted bypass is more relevant during slower deformation at elevated temperatures. Thermally activated dislocation climb is not expected to contribute significantly over the short duration of dislocation residence time, as discussed in Section~\ref{relaxation}. Therefore, the oxide particle contribution is described using the Orowan bowing mechanism (SI Appendix, Eq.~S29).\\

\begin{figure}[htbp]
\centering
\includegraphics[width=\textwidth]{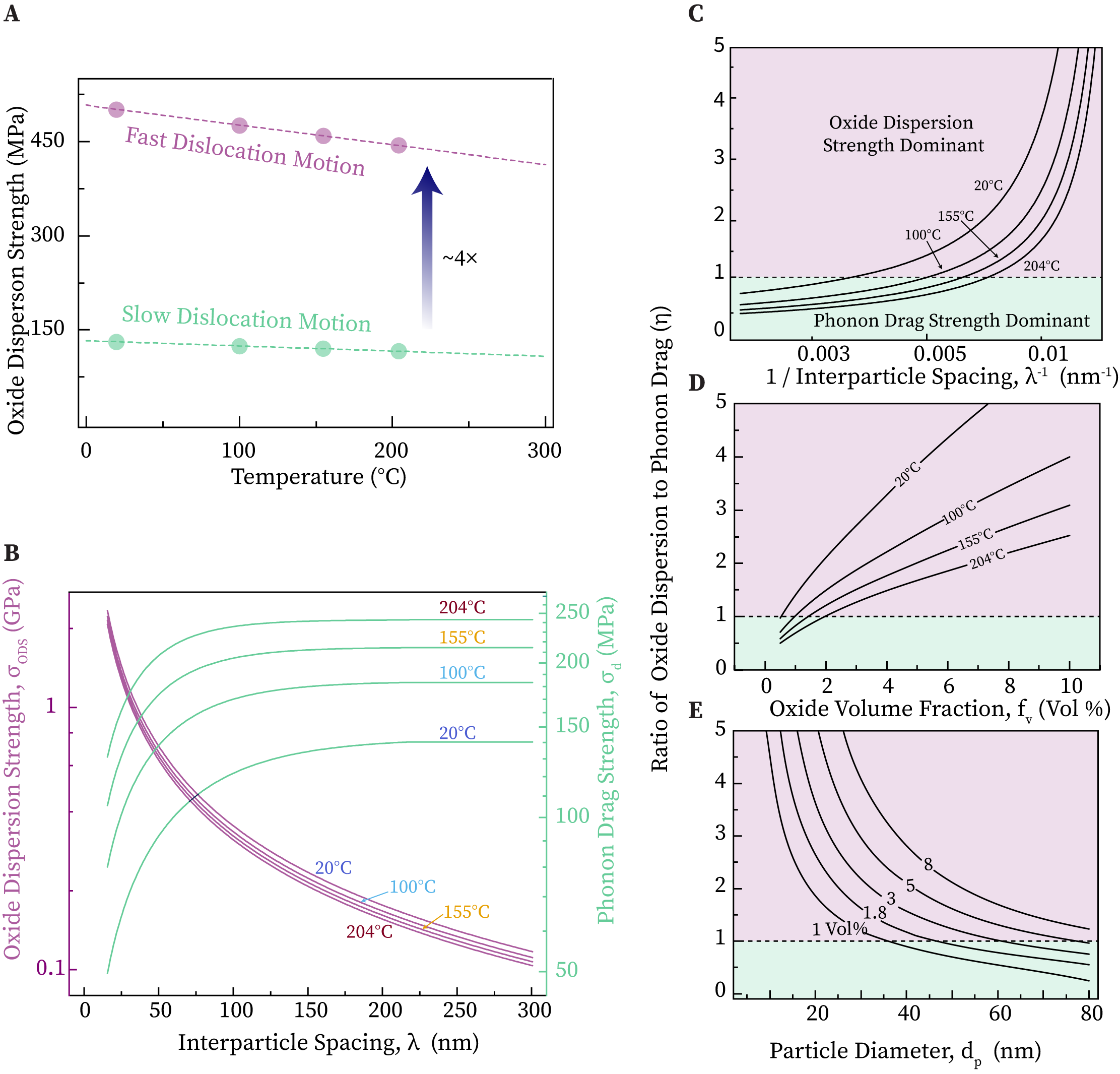}
\caption{\small\textbf{Competition between oxide dispersion and phonon drag strengthening.} (A) Oxide dispersion strength versus temperature, showing $\approx 4\times$ enhancement during fast dislocation motion. \textbf{(B)} Interplay between oxide dispersion strength and phonon drag strength across varying interparticle spacings at temperatures of $20~^\circ\text{C}$, $100~^\circ\text{C}$, $155~^\circ\text{C}$, and $204~^\circ\text{C}$. \textbf{(C)} Ratio of oxide dispersion strength to phonon drag strength ($\eta$) as a function of inverse interparticle spacing. The dashed line at a ratio of one separates oxide dispersion strength dominance, shown in purple, from phonon drag strength dominance, shown in green. \textbf{(D)} Ratio of oxide dispersion strength to phonon drag strength as a function of oxide volume fraction at each test temperature for a fixed particle diameter of 25 nm. \textbf{(E)} Ratio of oxide dispersion strength to phonon drag strength as a function of particle diameter for several oxide volume fractions at $20~^\circ\text{C}$.}
\label{figoxide_dispersion_phonon_drag}
\end{figure}

\noindent Oxide dispersion strengthening and phonon drag exhibit opposing dependencies on temperature and interparticle spacing, as shown in Fig.~\ref{figoxide_dispersion_phonon_drag}B. Phonon drag strength increases with temperature because the drag coefficient rises as lattice vibrations intensify. Although oxide dispersion strengthening is conventionally athermal, its magnitude decreases with temperature because of the reduction in elastic constants (SI Appendix, Eqs.~S3 and S4). Oxide dispersion strength also decreases with increasing interparticle spacing because the Orowan contribution is inversely related to the distance between particles. In contrast, larger spacings allow dislocations to accelerate over longer uninterrupted distances, increasing their interaction with lattice vibrations and strengthening the phonon drag contribution. Oxide dispersion strengthening dominates at smaller spacings, whereas phonon drag becomes increasingly important at larger spacings. The crossover position varies with temperature because both mechanisms depend on temperature sensitive parameters, including the elastic constants and the phonon drag coefficient. The transition between the two mechanisms is further illustrated in Fig.~\ref{figoxide_dispersion_phonon_drag}C. The strength ratio ($\eta$), defined as the oxide dispersion strengthening contribution divided by the phonon drag strengthening contribution, indicates oxide dispersion dominance when $\eta>1$ and phonon drag dominance when $\eta<1$. The nonlinear increase in this ratio with inverse interparticle spacing demonstrates the strong sensitivity of the governing mechanism to oxide spacing. Consequently, modest changes in spacing may shift the material between the two regimes. \\

\noindent Oxide volume fraction and particle diameter provide additional control over the above mentioned strength competition. As shown in Fig.~\ref{figoxide_dispersion_phonon_drag}D, the strength ratio increases with oxide volume fraction because a greater oxide population reduces the distance available for uninterrupted dislocation motion and increases the frequency of dislocation particle interactions. The separation between the temperature dependent curves also shows that oxide volume fraction must be considered together with the temperature dependent dislocation response. Figure~\ref{figoxide_dispersion_phonon_drag}E shows the combined influence of particle diameter and oxide volume fraction. At a fixed volume fraction, the strength ratio decreases with increasing particle diameter because larger particles reduce the particle number density and increase the effective obstacle spacing. The ratio of unity identifies combinations for which oxide dispersion strengthening and phonon drag contribute equally. Smaller particles and higher oxide volume fractions favor oxide dispersion dominance, whereas larger particles and lower volume fractions increase the relative contribution of phonon drag. These relationships provide a framework for tailoring the response of oxide particle containing superalloys during fast dislocation motion.

\section{Conclusion}

\noindent This work uncovers a key fundamental aspect of fast dislocation motion in oxide dispersion strengthened superalloys by demonstrating that dense oxide networks actively regulate dislocation acceleration and relaxation rather than merely acting as static physical obstacles. We used a synchronized pulsed laser microparticle impact platform coupled with high speed, nanosecond resolved imaging to probe dislocation dynamics under extreme rate deformation. This approach enabled us to decouple the underlying spatial and temporal mechanisms governing dislocation motion at velocities approaching the material's shear wave speed. Our framework informed by high throughput experiments demonstrates that fast moving dislocations bypass fine oxide obstacles via localized bowing rather than conventional thermally activated climb, thereby suppressing departure side dislocation relaxation due to dislocation residence time on picosecond timescales. Concurrently, the tight spatial confinement imposed by the dense oxide particle network limits phonon dislocation interactions and modulates the macroscopic phonon drag contribution to dynamic flow strength. These findings shed light on picosecond dislocation kinetics and macroscale dynamic flow strength, demonstrating that fine oxide dispersoids fundamentally alter the energy dissipation pathways within the vibrating lattice.

\newpage

\section*{Materials and Methods}
\addcontentsline{toc}{section}{Materials and Methods}

\subsection*{\textbf{Materials Preparation}}

\addcontentsline{toc}{subsection}{Materials Preparation}

\noindent The impact substrate was an oxide particle containing superalloy, GRX-810, containing a 1.8 vol.\% fraction of yttria (Y$_2$O$_3$) oxide particles. The material was fabricated by Elementum 3D using laser powder bed fusion (LPBF) additive manufacturing based on parameters from prior studies~\cite{smith20233d,smith2025mechanisms,heczko2026microstructural}. The material was sectioned into $9 \times 3 \times 3~\mathrm{mm}$ specimens for microparticle impact testing. To obtain a flat and smooth impact surface, the specimens were sequentially ground using 800-grit, 1200-grit, and 2400-grit SiC papers. Final mechanical polishing was performed using $1~\mu\mathrm{m}$, $0.05~\mu\mathrm{m}$, and $0.02~\mu\mathrm{m}$ polishing suspensions to obtain a mirror finish before testing. The polished surfaces were then cleaned with acetone and ethanol to remove residual contaminants. Monodisperse silica ($\mathrm{SiO_2}$) microparticles with an average diameter of $13.79 \pm 0.59~\mu\mathrm{m}$ served as rigid impactors for all experiments. Silica microparticles were purchased from microParticle GmbH (Berlin, Germany). Prior to testing, their size and uniformity were verified via optical microscopy.

\subsection*{\textbf{Electron Microscopy Characterization}}
\addcontentsline{toc}{subsection}{Electron Microscopy Characterization}

\noindent Transmission electron microscopy (TEM) specimens were prepared from a $25.4 \times 25.4 \times 12.7~\mathrm{mm}$ GRX-810 specimen. The sample was mechanically ground to a thickness of $130~\mu\mathrm{m}$ using 600-grit SiC polishing paper. Disc specimens with a diameter of $3~\mathrm{mm}$ were then punched using a disc punch. To achieve electron transparency, punched out discs were electropolished in a solution of 92\% methanol and 8\% nitric acid at $-30~^\circ\mathrm{C}$ and 22 V using a Fischione Model 110 automatic twin jet electropolisher. Scanning transmission electron microscopy (STEM) imaging was performed on a JEOL JEM-2800 microscope operated at 200 kV using a high angle annular dark field (HAADF) detector. Compositional analysis was carried out using dual silicon drift detector (SDD) energy dispersive X-ray spectroscopy (EDS) on the same microscope. The compositional spectroscopy data were collected using NORAN System 7 software. Crystallographic orientation analysis was conducted using electron backscatter diffraction (EBSD) with a Velocity\texttrademark\ Super CMOS camera mounted on an FEI Teneo scanning electron microscope (SEM). Data collection was carried out at an accelerating voltage of 20~kV. Post impact crater morphology was examined using laser confocal microscopy (LCM) and SEM. Three dimensional surface height maps were obtained using an Olympus LEXT OLS5000 confocal microscope, enabling non-contact topographic characterization of impact craters. High resolution imaging of craters was performed using an FEI Teneo SEM operated at an accelerating voltage of 5~kV.

\subsection*{\textbf{Microparticle Impact Testing}}
\addcontentsline{toc}{subsection}{Micro Particle Impact Testing}

\noindent The controlled microparticle impact tests were conducted using a laser-induced particle impact testing (LIPIT) system, based on well established prior techniques~\cite{lee2014dynamic,veysset2016dynamics,hassani2020material}. The LIPIT system used in this study comprises four main components: (1) launching laser,  (2) launch pad assembly, (3) target stage, and (4) in situ optical imaging system for resolving impact and rebound velocities. The first component, the launching laser, was a pulsed Nd:YAG laser operating at 532 nm, with a 6 ns pulse duration and 2–20 mJ pulse energy. The laser system is a Q-smart 850 system from Lumibird. This laser was focused onto the launch pad, where rapid ablation of a small volume of the chromium (Cr) film generated the impulse needed to propel individual microparticles toward the target.\\

\noindent The second component, the launch pad assembly, was designed to support both ambient and elevated temperature testing. The launch pad consists of a $160\ \mu\text{m}$ thick, 25~mm diameter round glass cover slip coated with a 200~nm chromium (Cr) film, together with a UV curable adhesive (NOA 61) acquired from Norland Products, USA, and a second $220\ \mu\text{m}$ thick, 25~mm diameter round glass cover slip. Similar configurations have been demonstrated to provide controlled, repeatable microparticle impact testing across ambient and elevated temperatures~\cite{reiser2023microparticle,dowding2024metals,dowding2026extreme}.  The silica microparticles were deposited onto the glass surface and distributed uniformly using lens cleaning paper with a small amount of ethanol.\\

\noindent The third component, the target stage, positioned the superalloy substrate opposite the launch pad and enabled careful alignment under a 10$\times$ microscope objective to ensure controlled, repeatable impacts. For elevated temperature testing, the substrate was heated using a ceramic heating stage and tested at $20\text{ }^\circ\text{C}$, $100\text{ }^\circ\text{C}$, $155\text{ }^\circ\text{C}$, and $204\text{ }^\circ\text{C}$. Multiple K-type thermocouples were used to monitor and verify the target temperature throughout the experiments. Under these conditions, individual silica particles were launched toward the substrate at controlled velocities ranging from 70 to 218~m/s. \\

\noindent The fourth component, the in situ optical imaging system, was leveraged to capture particle motion. Impact and rebound events were illuminated with a SI-LUX 640 quasi continuous wave laser (640~nm, $30\ \mu\text{s}$ pulse duration) and recorded with a SIMX16 high speed camera (Specialized Imaging, USA). This high speed camera is capable of capturing images with a frame rate of nearly one billion frames per second. In this work, it captured 16 sequential frames with an exposure time of 10 ns and interframe spacings ranging from 250 to 370 ns (corresponding to 2.7–4.0 million frames per second), enabling nanosecond resolved observation of particle trajectories and impact dynamics. Together, these four components allowed high fidelity measurements of impact and rebound velocities to quantify the dynamic deformation response of the target material.

\section*{Data, Materials, and Software Availability}
\addcontentsline{toc}{section}{Data, Materials, and Software Availability}

\noindent All study data are included in the article and/or SI Appendix.

\section*{Acknowledgments}
\addcontentsline{toc}{section}{Acknowledgments}

\noindent N.D. gratefully acknowledges support from the Newell Family Graduate School Endowed Scholarship and the Copper Hansen Fellowship through the College of Mines and Earth Sciences at the University of Utah. The authors acknowledge support from the University of Utah College of Science Faculty Start-Up Fund and the Department of Metallurgical Engineering. This work made use of the University of Utah Nanofab shared facilities, supported, in part, by the MRSEC Program of the NSF under Award No. DMR-1121252.

\section*{Author contributions}
\addcontentsline{toc}{section}{Author contributions}

\noindent S.E.P. designed research; N.D. performed research; S.E.P. and N.D. analyzed data; S.E.P. and N.D. wrote the paper.

\section*{Competing interests}

\noindent The authors declare that they have no known competing financial interests or personal
relationships that could have appeared to influence the work reported in this paper.

\newpage

\renewcommand{\refname}{References}
\addcontentsline{toc}{section}{References}

\bibliographystyle{pnas2009} 
\bibliography{references} 

\end{document}


\begin{flushleft}
\Large \textbf{Fine Oxide Dispersoids Modulate Phonon Drag and Dislocation Relaxation in Dynamically Deformed Superalloys}
\end{flushleft}

\begin{flushleft}
\normalsize
Naveen Dinujaya$^{1,2}$, 
Suhas Eswarappa Prameela$^{*,1,2,3}$
\end{flushleft}

\vspace{0.5em}

{\small
\noindent $^{1}$ Department of Materials Science and Engineering, University of Utah, Salt Lake City, UT, USA \\
\noindent $^{2}$ Department of Metallurgical Engineering, University of Utah, Salt Lake City, UT, USA \\
\noindent $^{3}$ Department of Mechanical Engineering, University of Utah, Salt Lake City, UT, USA \\

\noindent *Corresponding author: suhas.prameela\,@\,utah.edu

\vspace{2em}
\hrule

\vspace{0.5cm}

\noindent \textbf{This PDF file includes:} 

\noindent Supporting text

\noindent Tables S1 to S4

\noindent Figures S1 to S5

\noindent Supplementary Movies S1-S4.

\noindent SI References

\section*{Supporting Information}

\begin{enumerate}

\item \textbf{Material Properties of the Impactor (SiO$_2$) and Substrate (GRX-810)}.

\setcounter{table}{0}
\renewcommand{\thetable}{S\arabic{table}}

\begin{table}[H]
\centering
\begin{threeparttable}
\caption{\small Material properties of the substrate \cite{carpenter_grx810_nasa_2025, heczko2026microstructural}.}
\label{tabsubstrate_materials}
\begin{tabular}{lccc}
\toprule
\textbf{Property} & \textbf{Symbol} & \textbf{Unit} & \textbf{Value} \\
\midrule
Density & $\rho_s$ & $\mathrm{kg}/\mathrm{m}^{3}$ & $\sim 8440$ \\
Poisson's ratio & $\nu_2$ & -- & $\sim 0.31$ \\
\bottomrule
\end{tabular}
\end{threeparttable}
\end{table}

\begin{table}[H]
\centering
\begin{threeparttable}
\caption{\small Material properties of the impactor \cite{sun2022hot,pabst2013elastic}.}
\label{tabimpactor_materials}
\begin{tabular}{lccc}
\toprule
\textbf{Property} & \textbf{Symbol} & \textbf{Unit} & \textbf{Value} \\
\midrule
Density & $\rho_i$ & $\mathrm{kg}/\mathrm{m}^{3}$ & $\sim 2200$ \\
Elastic modulus & $E_1$ & $\mathrm{GPa}$ & $\sim 70$ \\
Poisson's ratio & $\nu_1$ & -- & $\sim 0.17$ \\
\bottomrule
\end{tabular}
\end{threeparttable}
\end{table}

\newpage

\item{\textbf {Interparticle Spacing, Oxide Volume Fraction and Grain Size Before Impact}}

\begin{figure}[H]
    \centering
    \includegraphics[width=\textwidth]{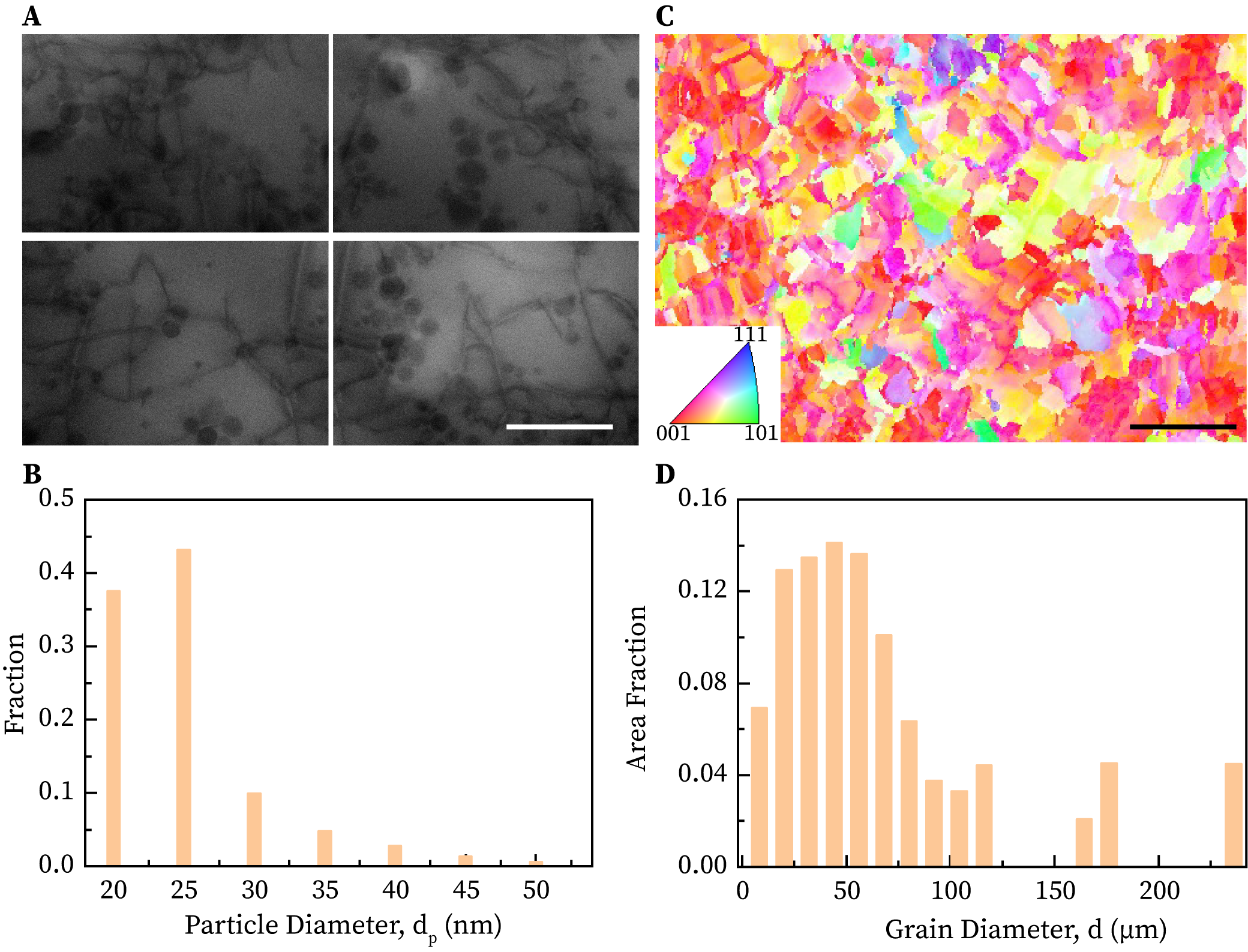}
    \caption{\small \textbf{Initial microstructure of GRX-810 before microparticle impact.} {(A)} STEM HAADF images showing nanoscale oxide dispersoids dispersed throughout the matrix. {(B)} Particle size distribution histogram indicating the fraction of dispersoid diameters ranging from $20\text{ to }50\text{ nm}$ with an average particle diameter of $25\text{ nm}$. 
    {(C)} Electron Backscatter Diffraction (EBSD) inverse pole figure (IPF) map displaying the grain structure and corresponding crystallographic orientation texture.  {(D)} Area fraction distribution histogram representing the grain diameter distribution with an average grain size of $40\ \mu\text{m}$ obtained from the EBSD map. Scale bars represent $250\text{ nm}$ (A) and $200\ \mu\text{m}$ (C).}
    \label{SI Fig 1}
\end{figure}

The average face to face interparticle spacing can be calculated using the equation below~\cite{zhang2009y2o3}.

\begin{equation}
\lambda = 1.25 \left( \sqrt[3]{1/N} \right) - 2r
\tag{S1}
\end{equation}

where $r = 12.5\text{ nm}$ is the average radius of the oxide particles, obtained from TEM image analysis indicating an average particle diameter of $25\text{ nm}$ (Fig.~\ref{SI Fig 1}A and B). The number density ($N$) can be determined with the below equation~\cite{zhang2009y2o3}.

\begin{equation}
N = \frac{f_v}{V_{\text{avg}}}
\tag{S2}
\end{equation}

where $f_v = 1.8\%$ is the volume fraction of the oxide particles and $V_{\text{avg}}$ is the average volume of an oxide particle. Additionally, EBSD analysis reveals an average initial grain size of $40\ \mu\text{m}$ (Fig.~\ref{SI Fig 1}C and D).

\item{\textbf{Temperature dependent Young's modulus and Shear Modulus}}
\label{Young's modulus and shear modulus}

\begin{table}[H]
\centering
\begin{threeparttable}
\caption{\small Temperature dependent Young's modulus values for
GRX-810 \cite{carpenter_grx810_nasa_2025}.}
\label{tabgrx810_elastic_modulus}
\begin{tabular}{cc}
\toprule
\textbf{Temperature ($^{\circ}\mathrm{C}$)}
&
\textbf{Young's modulus (GPa)}
\\
\midrule
$-196$ & 182.0 \\
21     & 158.6 \\
427    & 115.8 \\
649    & 106.2 \\
871    & 82.1  \\
1093   & 42.7  \\
\bottomrule
\end{tabular}
\end{threeparttable}
\end{table}

The temperature dependent Young's modulus was determined by a
least squares linear fit to the measured values listed in
Table~\ref{tabgrx810_elastic_modulus}. The fitted relation is

\begin{equation} E(T) = -0.101(T-273.15)+162.81 \tag{S3} \label{eqyoungs_modulus_temp} \end{equation} where $E(T)$ denotes the temperature dependent elastic modulus, expressed in GPa, and $T$ is the temperature in kelvin. The corresponding shear modulus is calculated considering isotropic elasticity, where $\nu$ is Poisson’s ratio \cite{laplanche2018elastic}. \begin{equation} G(T) = \frac{E(T)}{2(1+\nu)} \label{eqshear_modulus_temp} \tag{S4} \end{equation}

\item{\textbf{Strain Rate Calculation}}

\noindent In this work, strain rates below $10^{-3}~\mathrm{s}^{-1}$ are associated with slow dislocation motion, whereas strain rates above $10^{5}~\mathrm{s}^{-1}$ are associated with fast dislocation motion. The nominal strain rate during fast dislocation motion is estimated from prior microparticle impact testing studies~\cite{hassani2020material,li2026deformation}.

\begin{equation}
\dot{\varepsilon} = \frac{v_i}{d_i}
\tag{S5}
\end{equation}

\noindent where $d_i$ is the impactor diameter and $v_i$ is the impact velocity. The present experiments correspond to nominal strain rates of $5.07 \times 10^{6}$ to $1.58 \times 10^{7}~\mathrm{s}^{-1}$.

\begin{figure}[H]
    \centering
    \includegraphics[width=\textwidth]{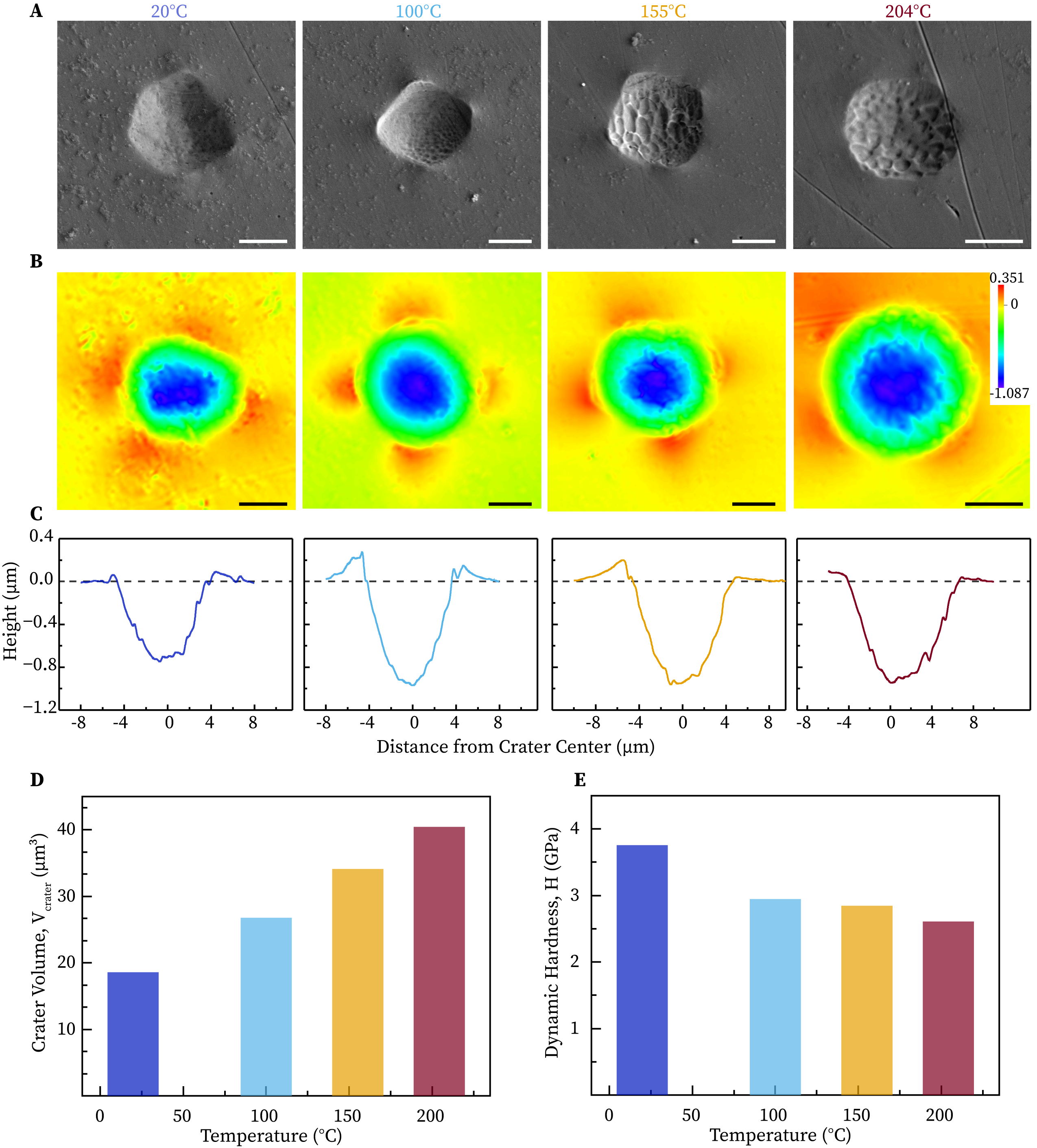}
    \caption{\small \textbf{Temperature dependent crater topography and derived mechanical response.} (A) Scanning electron microscopy (SEM) images showing surface morphology of indentation craters formed at 20\,\textdegree C, 100\,\textdegree C, 155\,\textdegree C, and 204\,\textdegree C. (B) Corresponding 2D laser confocal microscopy (LCM) maps displaying the craters (color bar indicates height in $\mu\text{m}$). (C) Cross sectional 1-D surface profiles extracted through the center of each crater as a function of distance. (D) Measured crater volume versus temperature, showing an increase in indentation volume with elevated temperature. (E) Material hardness as a function of temperature. Scale bars for (A) and (B) are $4~\mu\text{m}$.}
    \label{SI Fig 2}
\end{figure}

\item{\textbf{Dynamic Flow Strength from Microparticle Impact Testing}} 

\noindent The dynamic flow strength is evaluated by fitting the measured impact and rebound response using the Wu et al. ~\cite{wu2003rebound,wu2005energy} spherical impact model for elastic particles impacting elastic plastic substrates. This model has been extensively used in prior microparticle impact studies~\cite{liu2024model,dowding2024metals}. In our implementation, we measured the coefficient of restitution (CoR) and fitted the velocity data using the model's scaling relations to extract the dynamic flow strength ($Y_d$). For an elastic particle impacting an elastic-perfectly plastic substrate, the threshold impact velocity for the onset of plastic deformation ($v_y$) is expressed as 

\begin{equation}
v_y
=
\left(
\frac{26Y_d^5}{\rho_i E^{*4}}
\right)^{1/2}
\tag{S6}
\end{equation}

where $\rho_i$ is the density of the impactor. The effective modulus of the impactor substrate system, $E^{*}$, is given by

\begin{equation}
\frac{1}{E^{*}}
=
\frac{1-\nu_1^2}{E_1}
+
\frac{1-\nu_2^2}{E_2}
\tag{S7}
\end{equation}

where $E$ and $\nu$ are the Young's modulus and Poisson's ratio of the corresponding materials. Subscripts 1 and 2 denote the $\mathrm{SiO_2}$ impactor and the GRX-810 substrate, respectively.

\item{\textbf{Hardness Measurement During Fast Dislocation Motion}}

The temperature dependent crater morphology is shown by the SEM images in Fig.~\ref{SI Fig 2}A and the corresponding laser confocal microscopy 2D maps in Fig.~\ref{SI Fig 2}B. Cross sectional profiles through the crater centers are shown in Fig.~\ref{SI Fig 2}C. Using the measured crater volume, hardness during fast dislocation motion~\cite{hassani2020material,dowding2024metals,borasi2025self} is determined by normalizing the energy dissipated in the substrate by the microparticle impact crater volume.

\begin{equation}
  \text{Dynamic Hardness}~(H) = \frac{\frac{1}{2} m_p (v_i^2 - v_r^2)}{V_{\text{crater}}}
  \tag{S8}
\end{equation}

where $m_p$ is the impacting particle mass, $v_i$ and $v_r$ are the impact and rebound velocities, respectively, and $V_{\text{crater}}$ is the crater volume formed from microparticle impact testing. Crater volume increases with rising temperature, indicating enhanced plastic deformation (Fig.~\ref{SI Fig 2}D). The calculated hardness values (Fig.~\ref{SI Fig 2}E) exhibit a downward trend, consistent with the dynamic flow strength values obtained in the main paper Fig.~2D, highlighting the role of thermal weakening at elevated temperatures.

\item{\textbf{Flow Strength under Slow and Fast Dislocation Motion}}

\noindent Flow strength during slow and fast dislocation motion can be decomposed into several resistance contributions: matrix resistance~\cite{dowding2026extreme,de2019temperature}, short range barrier resistance~\cite{meyers1994dynamic,seils2020temperature}, athermal microstructural resistance~\cite{dowding2024metals,de2019temperature,seils2020temperature}, oxide particle resistance~\cite{scattergood1975orowan}, and dynamic dislocation resistance~\cite{dowding2024metals,dowding2026extreme}. Under slow dislocation motion, the flow strength is mainly controlled by solid solution strengthening, dislocation density strengthening, grain boundary strengthening, and oxide particle strengthening. However, under fast dislocation motion, these same microstructural barriers still contribute, but rate dependent effects become increasingly important~\cite{tang2024quantifying}. During fast dislocation motion, thermally activated resistance reflects the stress required to overcome short range barriers, and dynamic dislocation resistance arises from phonon drag during rapid dislocation glide~\cite{ashby1983mechanisms}. Consequently, the flow strength during fast dislocation motion includes both barrier controlled strengthening and velocity dependent drag resistance.

\noindent 7.1 \textbf{Slow Dislocation Motion} 

The flow strength of GRX-810 under slow dislocation motion depends on the following strengthening mechanisms~\cite{de2019temperature,seils2020temperature}:
i) solid solution strengthening $\left(\sigma_{ss}\right)$,
ii) Bailey-Hirsch dislocation density strengthening $\left(\sigma_{BH}\right)$,
iii) Hall-Petch grain boundary strengthening $\left(\sigma_{HP}\right)$, and
iv) oxide particle strengthening $\left(\sigma_K\right)$.
The total flow strength is taken as the sum of these individual strengthening contributions

\begin{equation}
\sigma_y^{SD}(T)
=
\sigma_{ss}(T)
+
\sigma_{BH}(T)
+
\sigma_{HP}(T)
+
\sigma_K(T)
\label{eqslow_total_strength}
\tag{S9}
\end{equation}

i) The solid solution strengthening contribution is calculated using the alloying element concentrations and strengthening coefficients. The temperature dependent strengthening coefficient for each solute is scaled with the shear modulus as

\begin{equation}
C_i(T)
=
C_{i,\mathrm{ref}}
\left(
\frac{G(T)}{G_{\mathrm{ref}}}
\right)
\label{eqsolute_coeff_temp}
\tag{S10}
\end{equation}

The strengthening coefficients used were
$C_{\mathrm{Co,ref}}=39.4~\mathrm{MPa}$,
$C_{\mathrm{Cr,ref}}=337~\mathrm{MPa}$,
$C_{\mathrm{W,ref}}=977~\mathrm{MPa}$, and
$C_{\mathrm{Re,ref}}=1000~\mathrm{MPa}$ \cite{xu2022modelling,li2023quantifying}.
The solute atomic fractions were
$x_{\mathrm{Co}}=0.31553$,
$x_{\mathrm{Cr}}=0.33527$,
$x_{\mathrm{W}}=0.00948$, and
$x_{\mathrm{Re}}=0.00468$ \cite{smith20233d}.
The total solid solution strengthening~\cite{li2023quantifying} is then evaluated as 

\begin{equation}
\sigma_{ss}(T)
=
\left[
x_{\mathrm{Co}}C_{\mathrm{Co}}^2(T)
+
x_{\mathrm{Cr}}C_{\mathrm{Cr}}^2(T)
+
x_{\mathrm{W}}C_{\mathrm{W}}^2(T)
+
x_{\mathrm{Re}}C_{\mathrm{Re}}^2(T)
\right]^{1/2}
\label{eqsolid_solution_strength}
\tag{S11}
\end{equation}

The resulting temperature dependent solid solution strengthening contribution is shown in Fig.~\ref{SI Fig 3}A.

ii) The Bailey--Hirsch strengthening contribution \cite{bailey1960dislocation} accounts for strengthening from the dislocation density and is calculated as

\begin{equation}
\sigma_{BH}(T)
=
M\alpha_{\mathrm{BH}} G(T)b\sqrt{\rho_{dis}}
\label{eqbailey_hirsch_strength}
\tag{S12}
\end{equation}

where $M=3.06$ is the Taylor factor, $\alpha_{\mathrm{BH}}=0.24$ is a constant parameter \cite{kocks1977theory},
$b=0.2528~\mathrm{nm}$ is the Burgers vector \cite{heczko2026microstructural}, and
$\rho_{dis}=1.5\times10^{14}~\mathrm{m^{-2}}$ is the dislocation density \cite{heczko2026microstructural}.

iii) The Hall--Petch contribution describes strengthening from grain boundaries~\cite{seils2020temperature}. The Hall-Petch coefficient is scaled with temperature using

\begin{equation}
K_{HP}(T)
=
K_{HP,\mathrm{ref}}
\left(
\frac{G(T)}{G_{\mathrm{ref}}}
\right)^{1/2}
\label{eqhall_petch_coeff_temp}
\tag{S13}
\end{equation}

where $K_{HP,\mathrm{ref}}=0.8~\mathrm{MPa}\sqrt{\mathrm{m}}$ is the Hall–Petch constant and for GRX-810, we adopt this Hall–Petch constant value, based on NiCoCr/CrCoNi-family literature \cite{rasooli2026searching,schneider2020analysis}. The Hall-Petch strengthening term is then calculated as

\begin{equation}
\sigma_{HP}(T)
=
K_{HP}(T)d^{-1/2}
\label{eqhall_petch_strength}
\tag{S14}
\end{equation}

where $d=40~\mathrm{\mu m}$ is the average grain size before the impact as depicted in Fig. \ref{SI Fig 1}. The corresponding dislocation density and grain boundary strengthening contributions are shown in Fig.~\ref{SI Fig 3}B.

iv) The oxide particle strengthening contribution is calculated using a dislocation detachment based expression \cite{rosler1990new}. A temperature dependent activation term is first defined as

\begin{equation}
\Theta(T)
=
\frac{
\ln\left(\frac{\dot{\varepsilon}_0}{\dot{\varepsilon}}\right)
k_BT
\left(\frac{b}{r}\right)
}{
G(T)b^3
}
\label{eqorowan_thermal_term}
\tag{S15}
\end{equation}

where $k_B=1.380649\times10^{-23}~\mathrm{J/K}$  is the Boltzmann constant,
$r=12.5~\mathrm{nm}$ is the particle radius,
$\dot{\varepsilon}_0=7.7\times10^8~\mathrm{s^{-1}}$ is the reference strain rate \cite{rosler1990new}, and
$\dot{\varepsilon}=1.0\times10^{-4}~\mathrm{s^{-1}}$ is the applied strain rate. The oxide particle strengthening contribution is then expressed as

\begin{equation}
\sigma_K(T)
=
G(T)\,0.84M
\left(\frac{b}{r}\right)
\left[
1-
\frac{\Theta^{2/3}(T)}{1-k}
\right]
\sqrt{\frac{3f_v}{2\pi}}
\sqrt{1-k^2}
\label{equpdated_orowan_strength}
\tag{S16}
\end{equation} 

Here, $k=0.77$ is the relaxation factor. The resulting oxide particle strengthening contribution is shown in Fig.~\ref{SI Fig 3}C. Combining the individual strengthening contributions according to Eq.~\ref{eqslow_total_strength} gives the predicted total flow strength, which is consistent with the experimental measurements in Fig.~\ref{SI Fig 3}D.

\noindent 7.2 \textbf{Relaxation Factor Calculation for Slow Dislocation Motion}

\noindent The relaxation factor (\(k\)) is calculated from the apparent stress exponent and apparent activation energy. The apparent activation energy is derived from the experimental creep data at a constant stress of $69\text{ MPa}$
\cite{semiatin2024constant}. First, the temperature dependent Young's modulus is converted to the shear modulus using Eqs. \ref{eqyoungs_modulus_temp} and \ref{eqshear_modulus_temp}. The shear modulus derivative (\(dG/dT\)) is evaluated over the same temperature range, which is discussed in section \ref{Young's modulus and shear modulus}. The normalized applied stress (\(\sigma/\sigma_{det}\))~\cite{rosler1990new} is then determined from

\begin{equation}
\frac{\sigma}{\sigma_{det}} = \left[ 1 + \frac{3 \left( Q_{\mathrm{app}} - Q_v \right)}{2 R T n_{\mathrm{app}} \left( 1 - \frac{T}{G} \frac{\mathrm{d}G}{\mathrm{d}T} \right)} \right]^{-1}
\tag{S17}
\end{equation}

The apparent activation energy, \(Q_{\mathrm{app}}=\left(\frac{\partial \ln \dot{\varepsilon}}{\partial T}\right)RT^{2}\), was determined to be \(345.05\,\mathrm{kJ\,mol^{-1}}\) \cite{semiatin2024constant}, \(Q_v\) is the activation energy for matrix volume diffusion taken as \(344.5\,\mathrm{kJ\,mol^{-1}}\) \cite{semiatin2024constant}, $n_{\mathrm{app}} = 7.2$ is the apparent stress exponent, extracted from creep data in the stress range of 21-62~\text{MPa} at 1093~$^\circ\text{C}$ \cite{smith2025mechanisms}, \(R\) is the universal gas constant, and \(T\) is the absolute temperature. The relaxation factor~\cite{rosler1990new} is subsequently calculated as

\begin{equation}
k
=
1-
\left[
\frac{2k_{\mathrm{B}}T}
{3Gb^2r}
\,
\frac{n_{\mathrm{app}}}
{\frac{\sigma}{\sigma_{det}}\sqrt{1-\frac{\sigma}{\sigma_{det}}}}
\right]^{2/3}
\tag{S18}
\end{equation}

A relaxation factor approaching unity indicates little or no relaxation of the dislocation line energy at the particle-matrix interface, whereas a lower value of \(k\) indicates greater interfacial relaxation and a stronger attractive particle dislocation interaction. 

\begin{figure}[H]
    \centering
    \includegraphics[width=\textwidth]{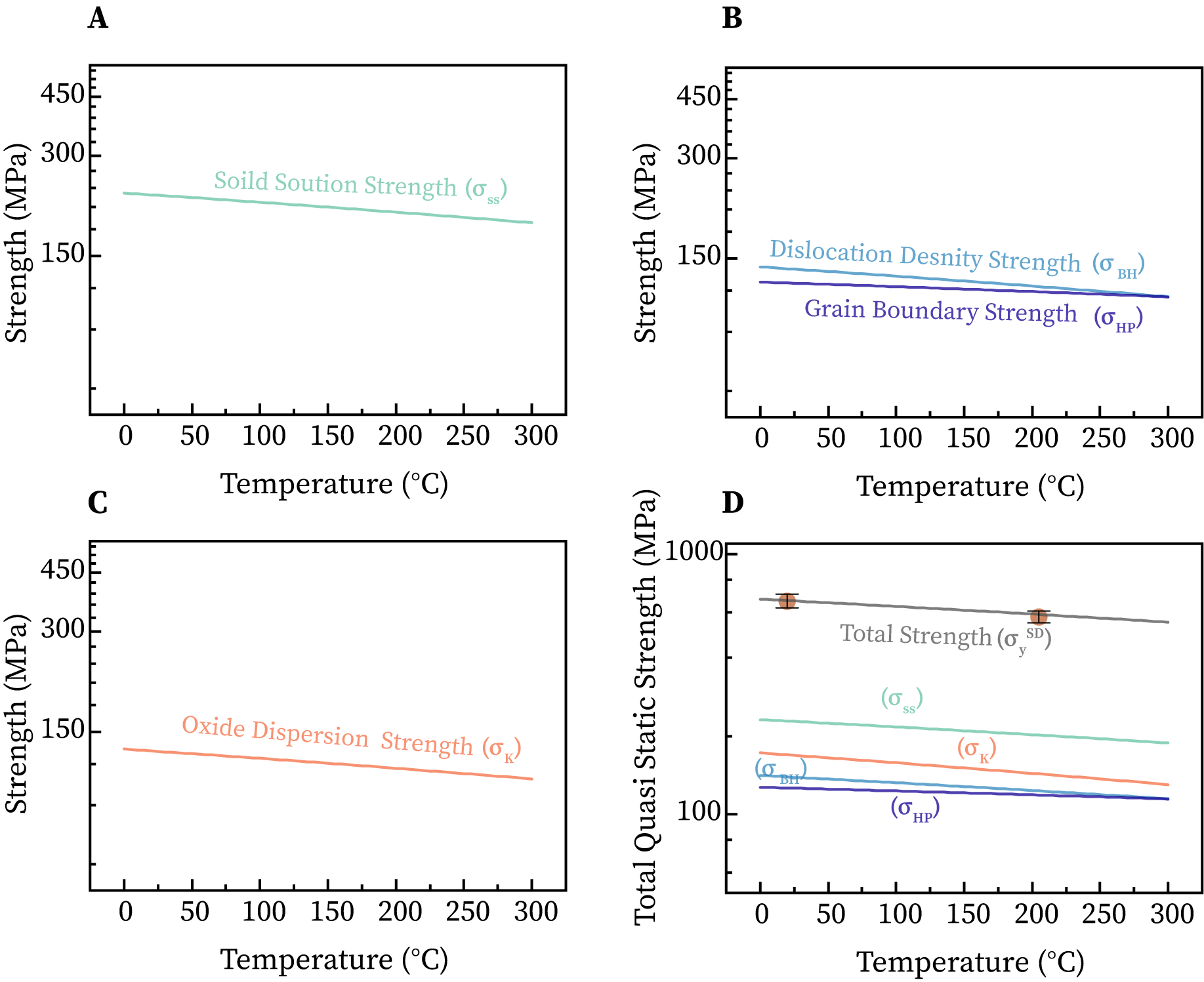}
    \caption{\small \textbf{Strength decomposition during slow dislocation motion.} {(A)} Solid solution strength, {(B)} Athermal resistance strength from dislocations and grain boundaries, {(C)} Oxide dispersion strength, {(D)} Total cumulative strength with experimental data. Experimental data for total cumulative strength values were extracted from available data \cite{carpenter_grx810_nasa_2025}.}
    \label{SI Fig 3}
\end{figure}

\noindent 7.3 \textbf{Fast Dislocation Motion}

The flow strength of GRX-810 during fast dislocation motion depends on several key factors. While previous microparticle impact studies have successfully decoupled standard baseline mechanisms \cite{tang2024quantifying,dowding2024metals,dowding2026extreme}, accurately modeling GRX-810 requires accounting for its defining fine oxide dispersoids. To address this, we expand upon the conventional framework by explicitly introducing a term for oxide particle resistance. The total flow strength is therefore formulated as the sum of distinct contributions: i) thermally activated short range barrier resistance $\left(\sigma_{th}\right)$,
ii) solute  resistance $\left(\sigma_s\right)$,
iii) athermal grain boundary resistance $\left(\sigma_{a,GB}\right)$,
iv) athermal dislocation density resistance $\left(\sigma_{a,disl}\right)$,
v) oxide particle resistance $\left(\sigma_{ODS}\right)$, and
vi) dynamic dislocation drag resistance $\left(\sigma_d\right)$.

\begin{equation}
\sigma_y^{FD}(T)
=
\sigma_{th}(T)
+
\sigma_s(T)
+
\sigma_{a,GB}(T)
+
\sigma_{a,disl}(T)
+
\sigma_{ODS}(T)
+
f_d\sigma_d(T).
\tag{S19}
\label{eqhigh_rate_total_strength}
\end{equation}

i) The thermally activated short range barrier resistance~\cite{meyers1994dynamic,dowding2024metals,dowding2026extreme} is calculated using

\begin{equation}
\sigma_{th}(T)
=
\sigma_0 \left[ 
1 - \left( \frac{k_B T}{g_0 G(T) b^3} \ln \left( \frac{\dot{\varepsilon}_0}{\dot{\varepsilon}} \right) \right)^{\frac{1}{q}} 
\right]^{\frac{1}{p}}
\tag{S20}
\label{eqXth_high_rate}
\end{equation}

where  $\sigma_0=689~\mathrm{MPa}$ is the threshold stress required to move a dislocation past the obstacle in the absence of thermal energy. For GRX-810, we use this value based on the reported 0 K critical resolved shear stress (CRSS) of 225 MPa for CrCoNi/NiCoCr, and convert it using the Taylor factor ($M=3.06$) to calculate stress \cite{li2022tensile}. T is the temperature, $g_0=0.8$ is a constant associated with the strength of the obstacles and $\dot{\varepsilon}_0=1.0\times10^{10}~\mathrm{s^{-1}}$ is the reference strain rate and
$\dot{\varepsilon}=1.0\times10^7~\mathrm{s^{-1}}$ is the applied strain rate. 
The values of $q=2/3$ and $p=2$, and $g_0$ are adopted from pure Ni, since GRX-810 is expected to exhibit FCC like matrix and dislocation behavior \cite{follansbee1990low}. The resulting temperature dependent short range barrier resistance is shown in Fig.~\ref{SI Fig 4}A.

ii) The solute resistance is calculated using a framework developed by Dowding and Schuh~\cite{dowding2026extreme}, for microparticle impact based analysis of fast dislocation motion for dilute solute in Ni alloys. We modified it to account for the principal substitutional solutes Co, Cr, W, and Re, while $\mathrm{Y_2O_3}$ is treated separately via the oxide dispersion strengthening term. Similarly, C, Nb, and Ti are excluded because their effects are primarily associated with carbide formation and particle–matrix interfacial strengthening~\cite{smith20233d,ma2025review}, rather than conventional matrix solid solution strengthening.

\begin{equation}
\sigma_{s}(T) = \left( \sum_{i} \sigma_{ci} \cdot x_i^{m} \right) \Bigg[ 1 - \left( \frac{k_B T}{g_0 G(T) b^3} \ln \left( \frac{\dot{\varepsilon}_0}{\dot{\varepsilon}} \right) \right) \Bigg]^{\frac{1}{q}}
\tag{S21}
\label{eqsigma_s_high_rate}
\end{equation}

where $\sigma_{ci}=0.22~\mathrm{GPa}$ and i = Co, Cr, W, Re. $\sigma_{ci}$ is the critical stress associated with each substitutional solute. In this analysis, an average value of 0.22 GPa is used for Co, Cr, W, and Re. The exponent $m=1.4$ describes the concentration dependence of solute spacing along the dislocation path and is obtained from the microparticle impact based model for FCC alloys \cite{dowding2026extreme}. $x_{\mathrm{i}}$ is the atomic fraction of solute ${\mathrm{i}}$ for each solute considered here. The corresponding solute resistance is shown in Fig.~\ref{SI Fig 4}B.

iii) The athermal microstructural resistance is separated into grain boundary and dislocation density contributions and modeled by adopting and modifying previous microparticle impact based studies~\cite{tang2024quantifying,dowding2026extreme}.

\begin{equation}
\sigma_a(T)
=
\sigma_{a,GB}(T)
+
\sigma_{a,disl}(T)
\tag{S22}
\label{eqathermal_total_high_rate}
\end{equation}

The grain boundary contribution is calculated as

\begin{equation}
\sigma_{a,GB}(T)
=
\left(\frac{K_{HP}(T)}{A} \right) G(T) \sqrt{\frac{b}{D}}
\tag{S23}
\label{eqsigma_a_GB_high_rate}
\end{equation}

where $A=5~\mathrm{MPa}\sqrt{\mathrm{m}}$ is a constant and $D=200~\mathrm{nm}$ is the grain size after impact, which
is assumed based on prior microparticle impact studies \cite{tiamiyu2022nanotwinning,cai2023dynamic,xie2017dynamics}. The dislocation density contribution is calculated as

\begin{figure}[H]
    \centering
    \includegraphics[width=\textwidth]{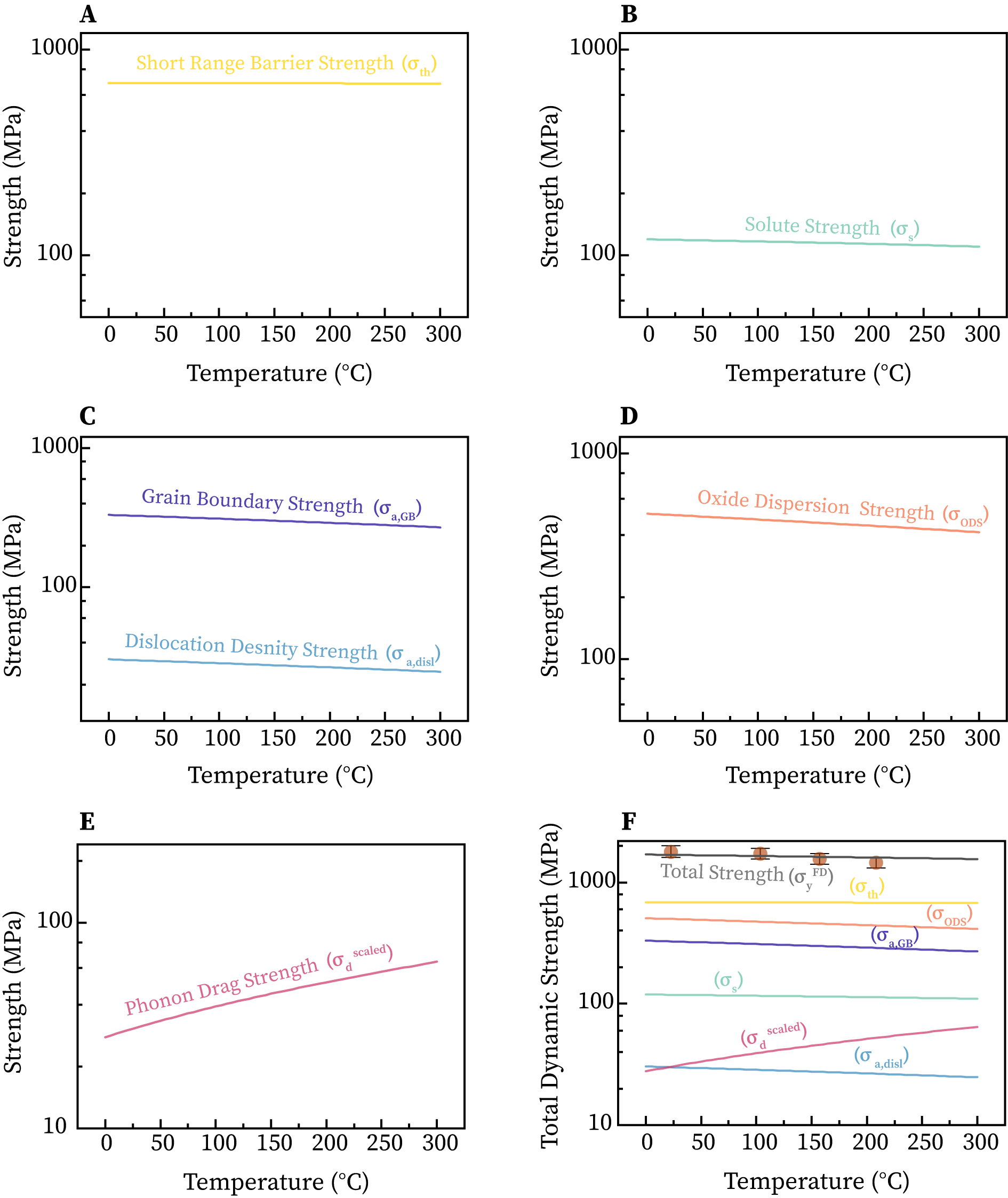}
  \caption{\small\textbf{{Strength decomposition during fast dislocation motion.}}
  {(A)} Short range barrier strength, {(B)} Solute strength, {(C)} Athermal resistance strength from dislocations and grain boundaries, {(D)} Oxide dispersion strength, {(E)} Phonon drag strength, {(F)} Total cumulative strength with experimental data.
}
\label{SI Fig 4}
\end{figure}

\begin{equation}
\sigma_{a,disl}(T)
=
\frac{M\alpha_{\mathrm{BH}} G(T)b\sqrt{\rho_m}}{B_{\rho}}
\tag{S24}
\label{eqsigma_a_disl_high_rate}
\end{equation}

where $B_{\rho}=1.46$ is a constant and $\rho_m=1.5\times10^{13}~\mathrm{m^{-2}}$ is the mobile dislocation density, which is estimated from the as built dislocation 
density. As built dislocation density is expected to be relatively high because of the dense 
oxide particle network produced by the additive manufacturing process \cite{heczko2026microstructural}. Under the 
present fast moving dislocation and temperature conditions, approximately 10\% of the 
total dislocation density is assumed to be mobile, consistent with fast moving dislocation plasticity models~\cite{tang2024quantifying, gasca1974model}. The combined athermal resistance from grain boundaries and mobile dislocations is shown in Fig.~\ref{SI Fig 4}C.

iv) Dynamic dislocation resistance is calculated by relating dislocation velocity to the imposed strain rate through the mobile dislocation density. The phonon drag contribution is written as a rate dependent stress term, modifying equations established in previous microparticle impact based studies~\cite{dowding2024metals,dowding2026extreme}.

\begin{equation}
B_d(T)
=
\left(
\frac{3k_BT}{5c(T)a^3}
\right)
(1.8b)
\tag{S25}
\label{eqdrag_coefficient_high_rate}
\end{equation}

where $B_d(T)$ is the drag coefficient, $c(T)$ is the temperature dependent shear wave velocity, and $a= 3.5751\times10^{-10}\ m$ is the lattice parameter. The shear wave velocity $c(T)$ is calculated from the temperature dependent shear modulus $G(T)$ and the mass density $\rho_{\text{s}}$ of the substrate.

\begin{equation}
    c{(T)} = \sqrt{\frac{G(T)}{\rho_{\text{s}}}}
    \tag{S26}
    \label{shear wave}
\end{equation}

Using this expression, the value of $B_d(T)$ is evaluated for each temperature, giving a temperature dependent characteristic drag coefficient for the GRX-810 matrix. The full phonon drag strength~\cite{dowding2024metals} is then evaluated as

\begin{equation}
\sigma_d(T)
=
\frac{B_d(T)\dot{\varepsilon}}
{\rho_{\mathrm{m}}b^2}
\tag{S27}
\label{eqdrag_strength_high_rate}
\end{equation}

In the present model, the drag contribution to the total strength is scaled using a factor obtained by fitting the strength model to the experimentally measured dynamic flow strengths.

\begin{equation}
\sigma_d^{scaled}(T)
=
f_d\sigma_d(T)
\tag{S28}
\label{eqscaled_drag_high_rate}
\end{equation}

where the fitted scaling factor $f_d\approx0.32$ corresponds to an approximately 68\% reduction relative to the unscaled phonon drag prediction. The possible role of dislocation confinement was subsequently assessed by comparing the interparticle spacing ($\lambda\approx70~\mathrm{nm}$) with the glide distance required to reach $99.5\%$ of the asymptotically approached steady state dislocation velocity ($\gamma\approx219~\mathrm{nm}$). The resulting phonon drag contribution is shown in Fig.~\ref{SI Fig 4}E.

v) The oxide particle resistance is calculated using an Orowan type expression~\cite{LLOYD19861745,chen2020situ}.

\begin{equation}
\sigma_{ODS}(T)
=
\frac{
0.84MG(T)b
}{
2\pi\sqrt{1-\nu}\lambda
}
\ln\left(
\frac{2r}{r_0}
\right)
\tag{S29}
\label{eqods_strength_high_rate}
\end{equation}

where $r_0=b$ is the dislocation core radius. The resulting oxide particle resistance is shown in Fig.~\ref{SI Fig 4}D. Combining the individual strengthening contributions according to Eq.~\ref{eqhigh_rate_total_strength} gives the predicted total flow strength, which is consistent with the experimentally measured dynamic flow strength in Fig.~\ref{SI Fig 4}F.

\item{\textbf{Applied Stress on Individual Dislocation During Fast Dislocation Motion}}

\noindent The applied stress $(\sigma_{\text{AP}})$ on individual dislocations is estimated using the dislocation acceleration framework developed by Tang et al.~\cite{tang2025suppressed} for microparticle impact based analysis of dynamic deformation. Their approach uses the portion of the measured impact strength that remains available to drive dislocation motion after accounting for microstructural barriers. We adopt this framework for GRX-810 and modify it to include the dominant strengthening terms relevant to this alloy. Specifically, $\sigma_{\mathrm{AP}}$ is obtained from the impact derived effective flow strength during fast dislocation motion ($Y_d$) after subtracting the intrinsic short range threshold resistance ($\sigma_0$), the Hall-Petch contribution before impact ($\sigma_{\mathrm{HP}}$) and the oxide dispersion contribution ($\sigma_{\mathrm{ODS}}$). These subtracted terms represent the main pre-existing barriers that limit the stress available for accelerating a dislocation segment through the matrix channel. The velocity dependent drag term is   not removed in this step because it is treated separately in the dislocation equation of motion. Thus, $\sigma_{\mathrm{AP}}$ should be viewed as an effective driving stress for the dislocation acceleration calculation, rather than as a complete or unique partitioning of the measured dynamic flow strength.

\begin{equation}
\sigma_{\mathrm{AP}}=Y_d-\sigma_0-\sigma_{\mathrm{HP}}-\sigma_{\mathrm{ODS}}
\tag{S30}
\label{single dislocation stress}
\end{equation}

\noindent The Hall-Petch contribution is calculated using Eq. \ref{eqhall_petch_strength}. This procedure gives an applied stress of $\sigma_{\mathrm{AP}}=\SI{447.06}{MPa}$ at $20~^\circ\mathrm{C}$, corresponding to an applied shear stress of $\tau_{\mathrm{AP}}=\SI{146.10}{MPa}$ after conversion using the Taylor factor ($M=3.06$).

\item{\textbf{Dislocation Acceleration Normalization and Drag Participation Factor}}

\noindent The dislocation acceleration under the applied shear stress is expressed as follows~\cite{gillis1970dislocation,tang2025suppressed}.

\begin{equation}
\frac{dv}{dt}
=
\frac{\tau_{\mathrm{AP}}\Gamma^{3/2}}{\rho_s b}
-
\frac{B_d(T)v\Gamma^{1/2}}{b^2\rho_s},
\qquad
\Gamma=1-\frac{v^2}{c^2}     
\tag{S31}
\label{drag acceleration equation}
\end{equation}

\noindent The dislocation velocity and acceleration are normalized as

\begin{equation}
v^*=\frac{v}{c},
\qquad
a^*=\frac{\rho_s b}{\tau_{\mathrm{AP}}}\frac{dv}{dt},
\qquad
\kappa=\frac{B_d(T)c}{\tau_{\mathrm{AP}}b}
\tag{S32}
\end{equation}

\noindent where $v^*$ is the normalized dislocation velocity, $a^*$ is the normalized acceleration, and $\kappa$ is the dimensionless drag parameter. Substitution of these terms into Eq.~\ref{drag acceleration equation} and factoring out the common velocity dependent term gives

\begin{equation}
a^*
=
\sqrt{1-(v^*)^2}
\left[1-(v^*)^2-\kappa v^*\right]
\tag{S33}
\end{equation}

\noindent The reduced normalized driving stress and drag resistance terms are therefore defined as $D^*(v^*) = 1-(v^*)^2$ and $R^*(v^*) = \kappa v^*$, respectively. The drag participation factor is defined as the ratio of these two terms

\begin{equation}
\Phi_D
=
\frac{R^*(v^*)}{D^*(v^*)}
=
\frac{\kappa v^*}{1-(v^*)^2}
=
\frac{B_d(T)v}{\tau_{\mathrm{AP}}b\Gamma}
\tag{S34}
\end{equation}

\noindent The drag participation factor represents the ratio of the normalized drag resistance to the normalized driving stress. A value of $\Phi_D=1$ represents an exact balance between the driving and drag contributions.

\begin{figure}[htbp]
    \centering
    \includegraphics[width=\linewidth]{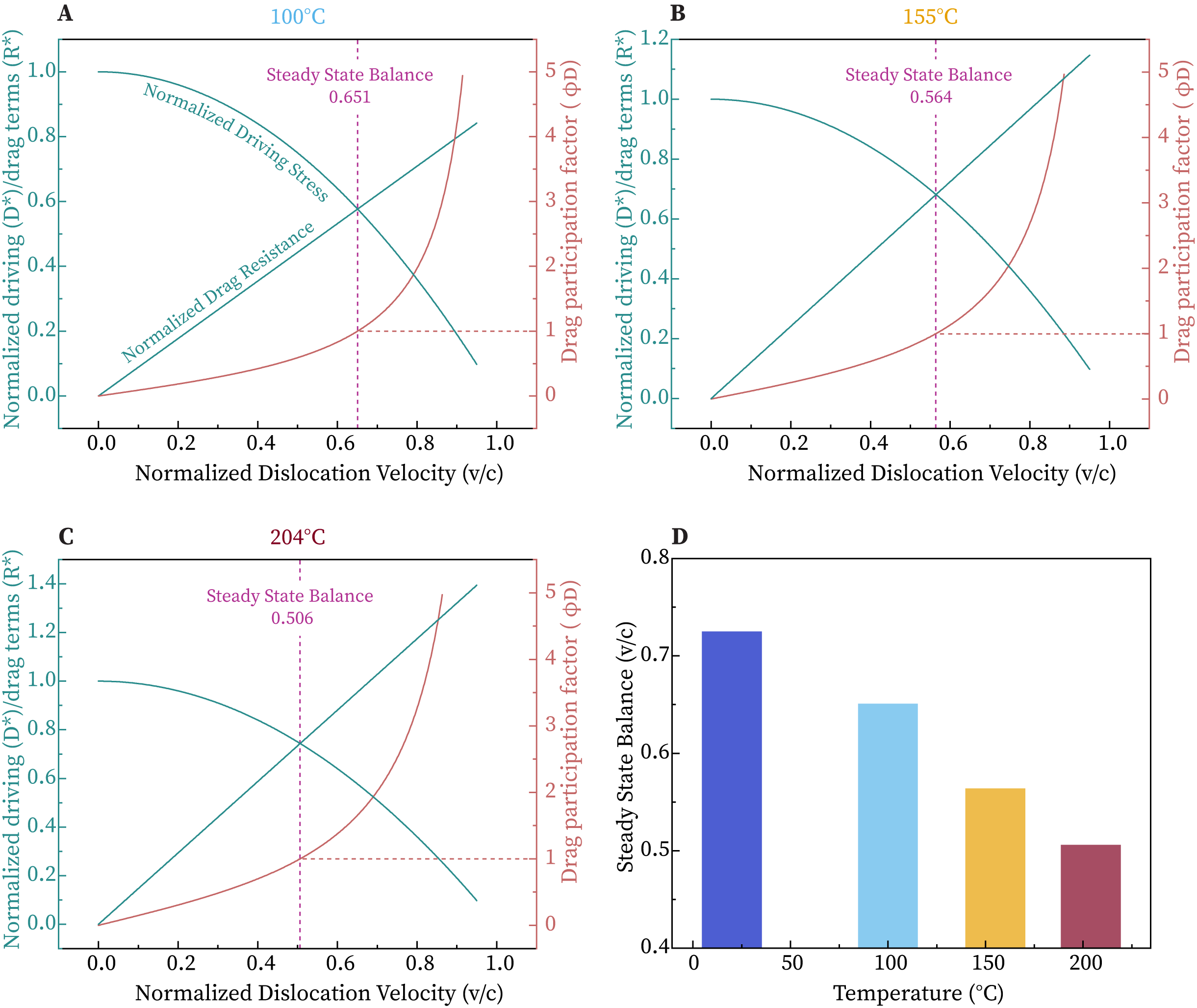}
    \caption{\textbf{Steady state dislocation kinetics across temperatures.}
    Normalized driving stress ($D^*$) and drag resistance terms ($R^*$) as a function of normalized dislocation velocity ($v/c$), alongside the drag participation factor ($\Phi_D$), expanding on main text Fig.~3D to elevated temperatures of (A) $100~^{\circ}\mathrm{C}$ ($v/c = 0.651$), (B) $155~^{\circ}\mathrm{C}$ ($v/c = 0.564$), and (C) $204~^{\circ}\mathrm{C}$ ($v/c = 0.506$). 
  (D) Steady state normalized dislocation velocity balance as a function of temperature.}
    \label{fig:SI_drag_participation}
\end{figure}

\noindent At this steady state balance condition, the normalized driving stress equals the drag resistance, establishing the steady state velocity, $v_{\mathrm{ss}}$. Extending this kinematic balance across elevated temperatures reveals a progressive reduction in the steady state velocity as depicted in Fig.~\ref{fig:SI_drag_participation}A-C. Specifically, the steady state balance point shifts from $v_{\mathrm{ss}}=0.725c$ at $20~^{\circ}\mathrm{C}$ to $0.651c$ at $100~^{\circ}\mathrm{C}$, $0.564c$ at $155~^{\circ}\mathrm{C}$, and $0.506c$ at $204~^{\circ}\mathrm{C}$ (Fig.~\ref{fig:SI_drag_participation}D).

\item{\textbf{Distance Based Analysis of Dislocation Acceleration}}

\noindent To determine how dislocation velocity and acceleration evolve with glide distance, the acceleration calculated from Eq.~\ref{drag acceleration equation} was combined with the kinematic relation $dx/dt=v$, giving

\begin{equation}
\frac{dx}{dv}
=
\frac{dx/dt}{dv/dt}
=
\frac{v}{dv/dt}
\label{eq:distance_velocity}
\tag{S35}
\end{equation}

\noindent Equation~\ref{eq:distance_velocity} was numerically integrated using the cumulative trapezoidal rule~\cite{matlab_cumtrapz} to obtain distance as a function of velocity, $x(v)$. The calculation used 20,000 uniformly spaced velocity points from $v_0=1~\mathrm{m/s}$ to $0.999~v_{\mathrm{ss}}$, with $x(v_0)=0$. The calculated velocity and acceleration were then plotted against the corresponding distance. Because the steady state velocity is approached asymptotically, 99.5\% of $v_{\mathrm{ss}}$ was selected as the numerical convergence criterion. The corresponding glide distance ($\gamma$) was estimated using the velocity grid point closest to $0.995~v_{\mathrm{ss}}$. For the present matrix properties and applied shear stress at $20~^\circ\mathrm{C}$, this calculation gives $\gamma\approx219~\mathrm{nm}$.

\item{\textbf{Dislocation Residence and Bowing Time During Fast Dislocation Motion}}

The temperature dependent shear modulus was calculated using Eq.~\ref{eqshear_modulus_temp}. The corresponding shear wave velocity and phonon drag coefficient were evaluated using Eqs.~\ref{shear wave} and \ref{eqdrag_coefficient_high_rate}, respectively. The temperature dependent dynamic flow strength, \(Y_d(T)\), was obtained from the interpolated experimental microparticle impact data for \(20~\mathrm{to}~204\,^{\circ}\mathrm{C}\) and from the temperature dependent strength model above \(204\,^{\circ}\mathrm{C}\). The stress available for continuous dislocation acceleration, \(\sigma_{\mathrm{AP}}(T)\), was calculated using Eq.~\ref{single dislocation stress}. The corresponding applied shear stress, \(\tau_{\mathrm{AP}}(T)\), was obtained using the Taylor factor conversion defined previously. The continuous fast moving calculation was applied only when

\begin{equation}
\sigma_{\mathrm{AP}}(T)>0
\label{eqpositive_driving_stress}
\tag{S36}
\end{equation}

At each temperature, the dislocation acceleration equation given by
Eq.~\ref{drag acceleration equation} was integrated simultaneously with the kinematic relation

\begin{equation}
\frac{dx}{dt}=v
\label{eqdislocation_position}
\tag{S37}
\end{equation}

where \(x\) is the distance traveled by the dislocation and \(v\) is
its instantaneous velocity. Immediately following interruption by an
oxide particle, the initial conditions were

\begin{equation} x(0)=0 \qquad v(0)=0 \tag{S38} \label{eqdislocation_initial_conditions} \end{equation} 

The coupled acceleration and position equations were integrated numerically at each temperature using the corresponding temperature dependent driving stress and drag resistance. Since the dislocation accelerates over the finite distance between oxide particles, the bowing and residence times were obtained directly from the calculated trajectory $x(t,T)$. Assuming that dislocation bowing occurs over one half of the mean oxide interparticle spacing~\cite{haussler2001hvtem,kato2008dislocation,scattergood1975orowan}, the bowing time was defined by \begin{equation} x\left(t_{\mathrm{bow}},T\right) = \frac{\lambda}{2} \label{eqbowing_time_definition} \tag{S39} \end{equation} The upper bound residence time was calculated by assuming that the dislocation interacts with three sides of the hexagonal particles, which is roughly equal to twice the average oxide particle diameter. \begin{equation} x\left(t_{\mathrm{res}},T\right) = 2d_{\mathrm{p}} \label{residence_time} \tag{S40} \end{equation} where $d_{\mathrm{p}}$ is the average oxide particle diameter.

\item{\textbf{Characteristic Dislocation Climb Time}}

\noindent The characteristic time for diffusion controlled dislocation climb was estimated using the attractive particle--dislocation interaction framework of Arzt and Rösler~\cite{arzt1988kinetics}. To obtain a conservative lower bound, the local climb branch was applied over the characteristic climb distance; inclusion of the preceding general climb stage would only increase the total bypass time. The dislocation line energy at the particle--matrix interface is

\begin{equation}
T_{\mathrm{A}} = k T_{\mathrm{D}},
\qquad
T_{\mathrm{D}} = \frac{1}{2}G(T)b^2
\tag{S41}
\label{eq:attractive_line_energy}
\end{equation}
where $T_{\mathrm{D}}$ is the reference dislocation line energy and $T_{\mathrm{A}}$ is the reduced line energy at the attractive particle-matrix interface. The relaxation factor ($k$) is determined in Section 7.2. In the Rösler-Arzt geometry, the particle spacing is expressed as $2\Lambda$. Since $\lambda$ denotes the measured face to face spacing in this work, the corresponding half center to center spacing is

\begin{equation}
\Lambda = \frac{\lambda+d_p}{2}
\tag{S42}
\label{eq:arzt_spacing}
\end{equation}
For $\lambda = 70~\mathrm{nm}$ and $d_p = 25~\mathrm{nm}$, $\Lambda = 47.5~\mathrm{nm}$. The particle face inclination is approximated as $\beta = 45^\circ$. For lattice diffusion controlled climb, the kinetic coefficient is

\begin{equation}
C_{\mathrm{cl}}(T) = \frac{2\pi D_v(T)d_p}{k_B T b}
\tag{S43}
\label{eq:climb_kinetic_coefficient}
\end{equation}
where $D_v(T)$ is the matrix lattice diffusivity. The Ni matrix lattice diffusivity is approximated using the monovacancy self-diffusion relation~\cite{maier1976self}.

\begin{equation}
D_v(T) = D_0 \exp\left(-\frac{Q_{\mathrm{Ni}}}{RT}\right)
\tag{S44}
\label{eq:ni_lattice_diffusion}
\end{equation}
where $D_0 = 9.2\times10^{-5}~\mathrm{m^2\,s^{-1}}$ and $Q_{\mathrm{Ni}} = 277.9~\mathrm{kJ\,mol^{-1}}$.
The local climb velocity for an attractive particle--dislocation interaction~\cite{arzt1988kinetics} is calculated as

\begin{equation}
v_{\mathrm{cl}}^{\mathrm{loc}} = C_{\mathrm{cl}} \frac{4a_v}{d_p^2\sin^2\beta\cos^2\beta} \left[ \tau_{\mathrm{AP}}\Lambda b - T_{\mathrm{A}}\sin\beta \right]
\tag{S45}
\label{eq:local_climb_velocity}
\end{equation}
where $a_v = \Omega/b$, with $\Omega = a^3/4$ for the FCC matrix. Using $a = 3.5751\times10^{-10}~\mathrm{m}$ gives $a_v = 4.52\times10^{-20}~\mathrm{m^2}$. A conservative characteristic climb time is obtained by taking the climb distance as one particle radius,

\begin{equation}
t_{\mathrm{climb}}^{\mathrm{min}} = \frac{d_p/2}{v_{\mathrm{cl}}^{\mathrm{loc}}}
\tag{S46}
\label{eq:minimum_climb_time}
\end{equation}

At $204~^\circ\mathrm{C}$, the highest experimental temperature, $G = 54.28~\mathrm{GPa}$, $\tau_{\mathrm{AP}} = 106.06~\mathrm{MPa}$, and $D_v = 3.50\times10^{-35}~\mathrm{m^2/s}$. With $b = 0.2528~\mathrm{nm}$, $k = 0.77$, and the geometry defined above, Eq.~\ref{eq:local_climb_velocity} gives $v_{\mathrm{cl}}^{\mathrm{loc}} = 1.26\times10^{-24}~\mathrm{m/s}$, corresponding to
$t_{\mathrm{climb}}^{\mathrm{min}} = 9.9\times10^{15}~\mathrm{s}$. This extremely long climb time is expected at low homologous temperature, where limited point defect mobility strongly suppresses diffusion controlled dislocation climb~\cite{kassner2015fundamentals}. The resulting lower bound is approximately $1.6\times10^{26}$ times longer than the corresponding $61.2~\mathrm{ps}$ dislocation residence time. To evaluate whether climb becomes kinetically accessible at substantially higher temperature, the calculation was extended to the model predicted $1200~^\circ\mathrm{C}$ condition. At this temperature, $G = 15.88~\mathrm{GPa}$, $\tau_{\mathrm{AP}} = 36.22~\mathrm{MPa}$, and $D_v = 1.29\times10^{-14}~\mathrm{m^2/s}$, giving $v_{\mathrm{cl}}^{\mathrm{loc}} = 7.22\times10^{-5}~\mathrm{m/s}$ and $t_{\mathrm{climb}}^{\mathrm{min}} = 1.73\times10^{-4}~\mathrm{s}$. Even at this elevated model predicted temperature, the minimum climb time remains approximately $3.5\times10^{5}$ times longer than the corresponding $492.6~\mathrm{ps}$ residence time. Thus, attractive local climb remains kinetically inaccessible over the available dislocation--particle interaction time across both the experimental and model predicted temperature ranges.

\item{\textbf{Relaxation Factor During Fast Dislocation Motion}}

The relaxation factor during fast dislocation motion is evaluated by comparing
the time available for a dislocation to interact with an oxide
particle (residence time) with the characteristic time required for atomic
redistribution. Following Semiatin et al.~\cite{semiatin2024constant},
rhenium (Re) was treated as the rate limiting diffusing solute in
GRX-810. Re diffusion is vacancy mediated and represents the
effective diffusion rate of the slowest solute and is not treated as an isolated vacancy migration rate. The residence time was calculated using Eq.~\eqref{residence_time}, following the procedure described in Section~11 of the SI Appendix. Semiatin et al.~\cite{semiatin2024constant} used a representative Re
diffusivity of \(D_{\mathrm{ref}}=1.0\times10^{-16}\,
\mathrm{~\mathrm{m^2/s}}\) at \(T_{\mathrm{ref}}=1366~\mathrm{K}\).
Using the activation energy for Re diffusion in Ni ($Q_{\mathrm{Re}} = 412 \times 10^{3}\ \mathrm{J\ mol^{-1}}$) reported by Zeng et al.~\cite{zeng2009study}, the temperature dependent Re diffusivity is given by

\begin{equation}
{
D_{\mathrm{Re}}(T)
=
D_{\mathrm{ref}}
\exp\left[
-\frac{Q_{\mathrm{Re}}}{R}
\left(
\frac{1}{T}
-
\frac{1}{T_{\mathrm{ref}}}
\right)
\right]
}
\tag{S47}
\end{equation}

To produce the shortest possible diffusion controlled relaxation
time, the atomic redistribution distance was conservatively taken
as one atomic jump, \(\ell_r=b\). Using the three dimensional
diffusion relation, the minimum
relaxation time~\cite{einstein1956investigations} is

\begin{equation}
{
\tau_{\mathrm{rel}}(T)
=
\frac{b^2}
{6D_{\mathrm{Re}}(T)}
}
\tag{S48}
\end{equation}

The one jump assumption minimizes the relaxation time and
maximizes the calculated amount of relaxation. Any redistribution
distance greater than \(b\) would increase
\(\tau_{\mathrm{rel}}\) and move the relaxation factor closer to
unity. The dimensionless relaxation ratio is

\begin{equation}
{
\chi(T,\dot{\varepsilon})
=
\frac{t_{\mathrm{res}}}  
{\tau_{\mathrm{rel}}}
}
\tag{S49}
\end{equation}

A first order kinetic expression is adopted for the dynamic
relaxation factor

\begin{equation}
{
k_{\mathrm{dyn}}
=
k
+
\left(1-k\right)
\exp(-\chi)
}
\tag{S50}
\end{equation}

where \(k\) is the relaxation factor value associated
with slow dislocation motion. In this formulation,
\(k_{\mathrm{dyn}}=1\) represents negligible relaxation, whereas
\(k_{\mathrm{dyn}}=k\) represents complete dislocation
relaxation. Because \(0\leq k\leq1\), a conservative lower
bound that does not require a specified value of
\(k\) is

\begin{equation}
{
k_{\mathrm{dyn}}
\geq
\exp(-\chi)
}
\tag{S51}
\end{equation}

The effective dislocation-oxide interaction distance was taken as twice the average oxide particle diameter. The mobile dislocation density defined previously was used without further modification. At \(20\,^{\circ}\mathrm{C}\), the dislocation-oxide interaction time is $t_{\mathrm{res}}
={4.70\times10^{-11}\ \mathrm{s}}$. The corresponding minimum relaxation time is
$\tau_{\mathrm{rel}}
= 4.83\times10^{53}\ \mathrm{s}$. The dimensionless relaxation ratio is therefore

\begin{equation}
\begin{aligned}
\chi
&=
\frac{t_{\mathrm{res}}}  
{\tau_{\mathrm{rel}}}
= \frac{4.70\times10^{-11
 }}
{4.83\times10^{53}}
=
9.74 \times10^{-65}
\end{aligned}
\tag{S52}
\end{equation}

Consequently, the conservative lower bound for the dynamic relaxation factor is

\begin{equation}
{
k_{\mathrm{dyn}}(20\,^{\circ}\mathrm{C})
\geq
\exp(-9.74\times10^{-65})
\approx 1
}
\tag{S53}
\end{equation}

Thus, vacancy mediated Re redistribution is effectively absent
during the dislocation--oxide interaction at
\(20\,^{\circ}\mathrm{C}\). At \(1200\,^{\circ}\mathrm{C}\), the
dislocation-oxide interaction time is $t_{\mathrm{res}} = {4.93\times10^{-10}\ \mathrm{s}}$. The corresponding minimum relaxation time is
$\tau_{\mathrm{rel}} = 7.61\times10^{-6}\ \mathrm{s}$. The dimensionless relaxation ratio is

\begin{equation}
\begin{aligned}
\chi
&=\frac{t_{\mathrm{res}}}  
{\tau_{\mathrm{rel}}}
=
\frac{4.93\times10^{-10}}
{7.61\times10^{-6}}
=
6.47\times10^{-5}
\end{aligned}
\tag{S54}
\end{equation}

Therefore, the conservative lower bound for the relaxation factor
is

\begin{equation}  
{
\begin{aligned}
k_{\mathrm{dyn}}(1200\,^{\circ}\mathrm{C})
&\geq
\exp(-6.47\times10^{-5})\\
&=
0.999935
\end{aligned}
}
\tag{S55}
\end{equation}

Consequently, vacancy mediated relaxation time remains much longer than the dislocation-oxide
interaction time. Therefore, $k_{\mathrm{dyn}}$ is effectively unity at both \(20\,^{\circ}\mathrm{C}\) and
\(1200\,^{\circ}\mathrm{C}\) for the strain rate of \(1.0\times10^{7}\ \mathrm{s^{-1}}\).

\item{\textbf{High Speed Movies}}

High speed videography recorded a $637 \times 478\ \mu\text{m}$ field of view over $3.75$ to $5.55~\mu\mathrm{s}$, with playback slowed to $5\text{ frames per second}$.

\noindent \textbf{Movie S1.} High speed video showing silica microparticle impact on GRX-810 at 20\,\textdegree C with an incident velocity of 142\,m/s and a rebound velocity of 121\,m/s.\\[0.5em]
\noindent \textbf{Movie S2.} High speed video showing silica microparticle impact on GRX-810 at 100\,\textdegree C with an incident velocity of 142\,m/s and a rebound velocity of 118\,m/s.\\[0.5em]
\noindent \textbf{Movie S3.} High speed video showing silica microparticle impact on GRX-810 at 155\,\textdegree C with an incident velocity of 143\,m/s and a rebound velocity of 113\,m/s.\\[0.5em]
\noindent \textbf{Movie S4.} High speed video showing silica microparticle impact on GRX-810 at 204\,\textdegree C with an incident velocity of 143\,m/s and a rebound velocity of 110\,m/s.

\item{\textbf{Nomenclature}}
\end{enumerate}

\begingroup
\renewcommand{\thetable}{S4}
\begin{longtable}{@{}p{0.52\linewidth} p{0.24\linewidth} p{0.18\linewidth}@{}}
\caption{\centering \small Nomenclature used in the main text and SI Appendix.}
\label{tabnomenclature}\\

\toprule
\textbf{Name} & \textbf{Symbol} & \textbf{Unit} \\
\midrule
\endfirsthead

\multicolumn{3}{c}{\small \textbf{Table \thetable{} continued}}\\[0.5em]
\toprule
\textbf{Name} & \textbf{Symbol} & \textbf{Unit} \\
\midrule
\endhead

\midrule
\multicolumn{3}{r}{\small Continued on next page} \\
\endfoot

\bottomrule
\endlastfoot

Absolute temperature
& $T$
& $\mathrm{K}$ \\

Apparent activation energy
& $Q_{\mathrm{app}}$
& $\mathrm{J\,mol^{-1}}$ \\

Apparent stress exponent
& $n_{\mathrm{app}}$
& -- \\

Applied creep stress
& $\sigma$
& $\mathrm{MPa}$ \\

Applied shear stress for dislocation acceleration
& $\tau_{\mathrm{AP}}$
& $\mathrm{MPa}$ \\

Applied strain rate
& $\dot{\varepsilon}$
& $\mathrm{s^{-1}}$ \\

Applied stress for dislocation acceleration
& $\sigma_{\mathrm{AP}}$
& $\mathrm{MPa}$ \\

Athermal dislocation density resistance
& $\sigma_{a,\mathrm{disl}}$
& $\mathrm{MPa}$ \\

Athermal grain boundary resistance
& $\sigma_{a,\mathrm{GB}}$
& $\mathrm{MPa}$ \\

Athermal total resistance
& $\sigma_a$
& $\mathrm{MPa}$ \\

Atomic area parameter for dislocation climb
& $a_v$
& $\mathrm{m^2}$ \\

Atomic redistribution distance
& $\ell_r$
& $\mathrm{m}$ \\

Atomic volume of the FCC matrix
& $\Omega$
& $\mathrm{m^3}$ \\

Attractive particle--matrix dislocation line energy
& $T_{\mathrm{A}}$
& $\mathrm{N}$ \\

Assumed grain size after impact
& $D$
& $\mathrm{m}$ \\

Average grain size before impact
& $d$
& $\mathrm{m}$ \\

Average oxide particle diameter
& $d_p$
& $\mathrm{m}$ \\

Average oxide particle volume
& $V_{\mathrm{avg}}$
& $\mathrm{m^3}$ \\

Bailey-Hirsch strengthening contribution
& $\sigma_{\mathrm{BH}}$
& $\mathrm{MPa}$ \\

Boltzmann constant
& $k_B$
& $\mathrm{J\,K^{-1}}$ \\

Burgers vector magnitude
& $b$
& $\mathrm{m}$ \\

Climb kinetic coefficient
& $C_{\mathrm{cl}}(T)$
& $\mathrm{m\,N^{-1}\,s^{-1}}$ \\

Coefficient of restitution
& $\mathrm{CoR}$
& -- \\

Crater volume
& $V_{\mathrm{crater}}$
& $\mathrm{m^3}$ \\

Critical solute resistance
& $\sigma_{ci}$
& $\mathrm{GPa}$ \\

Detachment stress
& $\sigma_{\mathrm{det}}$
& $\mathrm{MPa}$ \\

Dimensionless relaxation ratio
& $\chi$
& -- \\

Dimensionless drag parameter
& $\kappa$
& -- \\

Dislocation acceleration
& $\mathrm{d}v/\mathrm{d}t$
& $\mathrm{m/s^2}$ \\

Dislocation bowing time
& $t_{\mathrm{bow}}$
& $\mathrm{s}$ \\

Dislocation core radius
& $r_0$
& $\mathrm{m}$ \\

Dislocation density scaling constant
& $B_{\rho}$
& -- \\

Dislocation interaction coefficient
& $\alpha_{\mathrm{BH}}$
& -- \\

Dislocation position
& $x$
& $\mathrm{m}$ \\

Dislocation residence time
& $t_{\mathrm{res}}$
& $\mathrm{s}$ \\

Dislocation velocity
& $v$
& $\mathrm{m/s}$ \\

Drag participation factor
& $\Phi_D$
& -- \\

Dynamic dislocation drag resistance
& $\sigma_d$
& $\mathrm{MPa}$ \\

Dynamic flow strength from impact fitting
& $Y_d$
& $\mathrm{MPa}$ \\

Dynamic relaxation factor
& $k_{\mathrm{dyn}}$
& -- \\

Effective elastic modulus of the impactor--substrate system
& $E^*$
& $\mathrm{GPa}$ \\

Elastic moduli of the impactor (1) and substrate (2)
& $E_1,\ E_2$
& $\mathrm{GPa}$ \\

Face to face interparticle spacing
& $\lambda$
& $\mathrm{m}$ \\

Flow strength during fast dislocation motion
& $\sigma_y^{\mathrm{FD}}$
& $\mathrm{MPa}$ \\

Flow strength during slow dislocation motion
& $\sigma_y^{\mathrm{SD}}$
& $\mathrm{MPa}$ \\

Grain boundary scaling constant
& $A$
& $\mathrm{MPa\,m^{1/2}}$ \\

Half center to center particle spacing
& $\Lambda$
& $\mathrm{m}$ \\

Hall-Petch strengthening contribution
& $\sigma_{\mathrm{HP}}$
& $\mathrm{MPa}$ \\

Hardness from impact crater analysis
& $H$
& $\mathrm{Pa}$ \\

Impact velocity
& $v_i$
& $\mathrm{m/s}$ \\

Impactor density
& $\rho_i$
& $\mathrm{kg\,m^{-3}}$ \\

Impactor diameter
& $d_i$
& $\mathrm{m}$ \\

Initial dislocation velocity for the distance integration
& $v_0$
& $\mathrm{m/s}$ \\

Lattice parameter
& $a$
& $\mathrm{m}$ \\

Local climb velocity
& $v_{\mathrm{cl}}^{\mathrm{loc}}$
& $\mathrm{m/s}$ \\

Matrix lattice diffusivity for dislocation climb
& $D_v(T)$
& $\mathrm{m^2/s}$ \\

Matrix volume diffusion activation energy
& $Q_v$
& $\mathrm{J\,mol^{-1}}$ \\

Microparticle mass
& $m_p$
& $\mathrm{kg}$ \\

Minimum characteristic climb time
& $t_{\mathrm{climb}}^{\mathrm{min}}$
& $\mathrm{s}$ \\

Minimum relaxation time
& $\tau_{\mathrm{rel}}$
& $\mathrm{s}$ \\

Mobile dislocation density
& $\rho_m$
& $\mathrm{m^{-2}}$ \\

Normalized dislocation acceleration
& $a^*$
& -- \\

Normalized applied stress
& $\sigma/\sigma_{\mathrm{det}}$
& -- \\

Normalized dislocation velocity
& $v^*=v/c$
& -- \\

Normalized obstacle strength parameter
& $g_0$
& -- \\

Ni lattice self-diffusion activation energy
& $Q_{\mathrm{Ni}}$
& $\mathrm{J\,mol^{-1}}$ \\

Ni lattice self-diffusion prefactor
& $D_0$
& $\mathrm{m^2\,s^{-1}}$ \\

Number density of oxide particles
& $N$
& $\mathrm{m^{-3}}$ \\

Oxide dispersion resistance during fast motion
& $\sigma_{\mathrm{ODS}}$
& $\mathrm{MPa}$ \\

Oxide particle radius
& $r$
& $\mathrm{m}$ \\

Oxide particle strengthening during slow motion
& $\sigma_K$
& $\mathrm{MPa}$ \\

Oxide particle volume fraction
& $f_v$
& -- \\

Oxide strengthening to phonon drag strength ratio
& $\eta$
& -- \\

Particle face inclination angle
& $\beta$
& $^\circ$ \\

Phonon drag coefficient
& $B_d(T)$
& $\mathrm{Pa\,s}$ \\

Phonon drag scaling factor
& $f_d$
& -- \\

Poisson's ratio
& $\nu$
& -- \\

Poisson's ratios of the impactor and substrate
& $\nu_1,\ \nu_2$
& -- \\

Re diffusion activation energy
& $Q_{\mathrm{Re}}$
& $\mathrm{J\,mol^{-1}}$ \\

Re diffusivity
& $D_{\mathrm{Re}}(T)$
& $\mathrm{m^2/s}$ \\

Rebound velocity
& $v_r$
& $\mathrm{m/s}$ \\

Reduced normalized drag term
& $R^*(v^*)$
& -- \\

Reduced normalized driving term
& $D^*(v^*)$
& -- \\

Reference dislocation line energy
& $T_{\mathrm{D}}$
& $\mathrm{N}$ \\

Reference Hall-Petch coefficient
& $K_{\mathrm{HP,ref}}$
& $\mathrm{MPa\,m^{1/2}}$ \\

Reference Re diffusivity
& $D_{\mathrm{ref}}$
& $\mathrm{m^2/s}$ \\

Reference shear modulus
& $G_{\mathrm{ref}}$
& $\mathrm{GPa}$ \\

Reference strain rate
& $\dot{\varepsilon}_0$
& $\mathrm{s^{-1}}$ \\

Reference strengthening coefficient of solute $i$
& $C_{i,\mathrm{ref}}$
& $\mathrm{MPa}$ \\

Reference temperature
& $T_{\mathrm{ref}}$
& $\mathrm{K}$ \\

Relaxation factor during slow motion
& $k$
& -- \\

Relativistic factor
& $\Gamma$
& -- \\

Scaled phonon drag resistance
& $\sigma_d^{\mathrm{scaled}}$
& $\mathrm{MPa}$ \\

Shear modulus
& $G$
& $\mathrm{GPa}$ \\

Shear modulus temperature derivative
& $\mathrm{d}G/\mathrm{d}T$
& $\mathrm{GPa\,K^{-1}}$ \\

Shear wave velocity
& $c$ or $c(T)$
& $\mathrm{m/s}$ \\

Short range thermally activated resistance
& $\sigma_{\mathrm{th}}$
& $\mathrm{MPa}$ \\

Solid solution strengthening contribution
& $\sigma_{\mathrm{ss}}$
& $\mathrm{MPa}$ \\

Solute atomic fraction
& $x_i$
& -- \\

Solute concentration exponent
& $m$
& -- \\

Solute resistance during fast motion
& $\sigma_s$
& $\mathrm{MPa}$ \\

Steady state dislocation velocity
& $v_{\mathrm{ss}}$
& $\mathrm{m/s}$ \\

Substrate density
& $\rho_s$
& $\mathrm{kg\,m^{-3}}$ \\

Taylor factor
& $M$
& -- \\

Temperature dependent Hall--Petch coefficient
& $K_{\mathrm{HP}}(T)$
& $\mathrm{MPa\,m^{1/2}}$ \\

Temperature dependent shear modulus
& $G(T)$
& $\mathrm{GPa}$ \\

Temperature dependent strengthening coefficient of solute $i$
& $C_i(T)$
& $\mathrm{MPa}$ \\

Temperature dependent Young's modulus
& $E(T)$
& $\mathrm{GPa}$ \\

Thermal activation exponents
& $p,\ q$
& -- \\

Thermal activation parameter for particle detachment
& $\Theta(T)$
& -- \\

Threshold impact velocity for plastic deformation
& $v_y$
& $\mathrm{m/s}$ \\

Threshold stress at zero thermal activation
& $\sigma_0$
& $\mathrm{MPa}$ \\

Time
& $t$
& $\mathrm{s}$ \\

Total dislocation density
& $\rho_{\mathrm{dis}}$
& $\mathrm{m^{-2}}$ \\

Uninterrupted dislocation glide distance
& $\gamma$
& $\mathrm{m}$ \\

Universal gas constant
& $R$
& $\mathrm{J\,mol^{-1}\,K^{-1}}$ \\

Wu plastic impact prefactor
& $\alpha$
& -- \\

Young's modulus
& $E$
& $\mathrm{GPa}$ \\

\end{longtable}
\endgroup

\renewcommand{\refname}{SI References}
\bibliographystyle{pnas2009} 
\bibliography{references}